\documentclass[a4paper,11pt]{article}
\pdfoutput=1 	
\usepackage{jheppub} 	
\usepackage{verbatim}
\usepackage{graphicx}	
\usepackage{dcolumn}	
\usepackage{bm}		    
\usepackage{hyperref}
\usepackage{amsmath}
\usepackage{mathtools}
\usepackage{tabularx}
\usepackage{xfrac}
\usepackage{amssymb}
\usepackage{physics}
\usepackage{tikz-feynman}
\usepackage{subcaption}
\usepackage{float}
\usepackage{longtable}

\newcommand{\hc}{\mathrm{h.c.}}
\newcommand{\lagr}{\mathcal{L}}
\newcommand{\exlab}{X}

\title{\boldmath  Classification of minimal models producing $b$, $c$ and $\tau$ masses at one-loop level}

\author{Lucia Stockdale}
\author{and Raymond R. Volkas}

\affiliation{
ARC Centre of Excellence for Dark Matter Particle Physics, School of Physics, The University of Melbourne, Victoria 3010, Australia
}
\emailAdd{lucia.stockdale@student.unimelb.edu.au}
\emailAdd{raymondv@unimelb.edu.au}

\abstract{
The Standard Model neither provides a dynamical explanation for the quark and lepton mass hierarchies, nor a rationale for why all of these masses save that of the top quark are suppressed compared to the electroweak scale. Motivated by this, we explore the alternative radiative mass generation hypothesis, specialising to the generation of the $b$, $c$ and $\tau$ masses at one-loop level. A classification of all minimal models that use exotic scalars and exotic fermions only is presented, resulting in $25$ possibilities featuring four exotic multiplets. As a bonus, some of the models produce a one-loop neutrino Dirac mass if a right-handed neutrino field is included, and some feature WIMP-like dark matter candidates. By way of example, we analyse the phenomenology of a benchmark model chosen from the set of $25$ candidates and find that it is capable of exactly replicating the predicted Standard Model Yukawa couplings within the permitted parameter space. 
}

\begin{document}
\maketitle
\flushbottom

\section{Introduction}

Apart from the top quark, quarks and leptons have masses that are hierarchically suppressed compared to the electroweak scale. Despite the technical naturalness of the Standard Model (SM) tree-level masses, the hierarchical distribution of the suppressed fermion masses has no dynamical origin in the SM. Because it can explain such a pattern, radiative mass generation for all charged fermions except the top quark is a compelling alternative to standard tree-level mass generation.\footnote{The extremely suppressed neutrino mass scale is an egregious and very important special case, but we concentrate on the charged fermions in this paper.} The fact that masses arise at loop level accounts for the generic fact of suppression. The hierarchy can then arise from masses that are radiatively generated at sequentially higher loop orders, a possibility that has been studied extensively~\cite{radiative1, radiative2, radiative3, radiative4, radiative5, radiative6, radiative7, radiative8, radiative9, radiative10, radiative14, radiative11, radiative12, radiative13, radiative15, Fraser_2014, Baker_2021, Stockdale:2025sxi}.\footnote{Radiative mass generation is not the unique possibility. For example, there is a large literature on using family symmetries and/or seesaw suppression; some representative papers are~\cite{Froggatt:1978nt,Davidson:1987mh,Rajpoot:1986nv,Davidson:1987tr,Davidson:1989bx,Davidson:1993xn,Hall:1995es,Pomarol:1995xc,Davidson:1998vr,King:2001uz,Carrasco-Martinez:2026wzu}.}

Any radiative model of course produces a panoply of testable beyond-SM (BSM) phenomenology. This includes BSM contributions to electroweak precision observables, collider constraints on the direct production of exotic particles, and, for some models, also dark matter candidates. Of particular note is that the SM predicts the relationship $y_f = \sqrt{2} m_f / v$, where $v$ is the SM vacuum expectation value (VEV), between fermion masses and associated Yukawa couplings. Models which modify how a fermion obtains its mass may also produce a corresponding modification to that fermion's Yukawa coupling, thus altering the mass-Yukawa relationship. Historically, modified Yukawa coupling constants have not been much analysed. They are, however, becoming increasingly important as the measurements of Higgs decay rates to fermions improve in precision. Multi-family radiative models are also expected to feature flavour-changing Higgs-fermion couplings, which will be subject to strict phenomenological constraints for the lower families.

One-loop radiative masses are well-motivated for the $b$ and $c$ quarks and the $\tau$ lepton, since their masses are roughly a factor of $1/16\pi^2$ below the electroweak scale $v$. In a recent work~\cite{Stockdale:2025sxi} we analysed a specific model generating the $b,c,\tau$ masses at one-loop level, as well as a one-loop Dirac mass for a single neutrino species. In this paper we generalise our previous effort by classifying all minimal models that generate one-loop effective Yukawas for $b,c,\tau$ using only additional scalars and vector-like fermions. We expect some of these will be able to form the basis of a complete model that treats all SM fermions at sequentially higher loop orders.

In Section~\ref{sec:model_class_section} we review the relevant background, describe the method by which we conducted the model classification, and present a catalogue of all possible minimal models. We examine the theoretical structure and phenomenology of a benchmark model chosen from this catalogue in Section~\ref{sec:h2_model}. We summarise and finish with some concluding remarks in Section~\ref{sec:conclusion}.

\section{Model classification}\label{sec:model_class_section}

We work in the framework formulated in~\cite{Baker_2021} and further developed in~\cite{Stockdale:2025sxi}. In Section~\ref{sec:notation} we briefly introduce the notation we adopt, before outlining the structure of this framework in Section~\ref{sec:BCV}. In Section~\ref{sec:minimal_models} we classify all minimal models within this framework that generate one-loop masses for all three of $b$, $c$, and $\tau$ simultaneously.

\subsection{Notation}\label{sec:notation}

\begin{figure}[t]
        \centering
        \begin{tikzpicture}
        \begin{feynman}
\vertex (a1);
\vertex[right=1.7cm of a1] (a2);
\vertex[right=2.3cm of a2] (a3);
\vertex[right=1.7cm of a3] (a4);
\vertex[right=1.15cm of a2] (fill);
\vertex[above=1.15cm of fill] (d);
\vertex[above=1.25cm of d] (e) {\(H\)};

\diagram* {
(a1) -- [fermion, edge label'=$f_R$] (a2) -- [ghost, with arrow =0.5, edge label'=\(\exlab_3^f\)] (a3) -- [fermion, edge label'=$F_L$] (a4),
(a2) -- [ghost, with arrow = 0.5, quarter left, edge label=\(\exlab_1^f\)] (d) -- [ghost, with arrow = 0.5, quarter left, edge label=\(\exlab_2^f\)] (a3),
(d) -- [scalar] (e),
};

        \end{feynman}
    \end{tikzpicture}
    \caption{Labelling convention adopted for the purpose of model building and classification. Arrows indicate the direction of particle flow. The various $\exlab_i$ may be either scalars or fermions.}
    \label{fig:excon}
\end{figure}
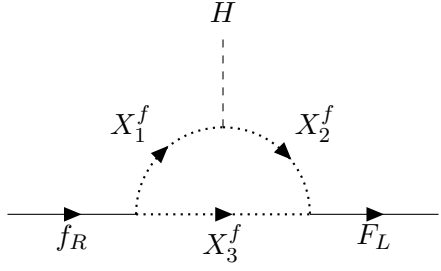

Let $f$ be an arbitrary SM fermion, with right-handed component $f_R$ and left-handed component contained in the $SU(2)_L$ doublet $F_L$. In our analysis we will consider $f \in \{b,c,\tau\}$. Since we will generate a Higgs Yukawa coupling for $f$ at one loop, it is useful to adopt the notation defined in Figure~\ref{fig:excon}, where the $\exlab_i$ are exotic fields which may be scalars or fermions. The exotic $\exlab_1^f$ couples to both $f_R$ and $H$, $\exlab_2^f$ to both $F_L$ and $H$, and $\exlab_3^f$ to both $f_R$ and $F_L$. 

\subsection{Model-building framework}\label{sec:BCV}

\begin{figure}
    \centering
    \begin{subfigure}[t]{0.49\textwidth}
        \centering
        \begin{tikzpicture}
        \begin{feynman}
\vertex (a1);
\vertex[right=1.7cm of a1] (a2);
\vertex[right=2.3cm of a2] (a3);
\vertex[right=1.7cm of a3] (a4);
\vertex[right=1.15cm of a2] (fill);
\vertex[above=1.15cm of fill] (d);
\vertex[above=1.25cm of d] (e) {\(H\)};

\diagram* {
(a1) -- [fermion, edge label'=$f_R$] (a2) -- [fermion] (a3) -- [fermion, edge label'=$F_L$] (a4),
(a2) -- [charged scalar, quarter left] (d) -- [charged scalar, quarter left] (a3),
(d) -- [scalar] (e),
};

\end{feynman}
        \end{tikzpicture}
        \caption{Class One}
        \label{fig:topologyc1}
    \end{subfigure}%
    ~ 
    \begin{subfigure}[t]{0.49\textwidth}
        \centering
        \begin{tikzpicture}
        \begin{feynman}
\vertex (a1);
\vertex[right=1.7cm of a1] (a2);
\vertex[right=2.3cm of a2] (a3);
\vertex[right=1.7cm of a3] (a4);
\vertex[right=1.15cm of a2] (fill);
\vertex[above=1.15cm of fill] (d);
\vertex[above=1.25cm of d] (e) {\(H\)};

\diagram* {
(a1) -- [fermion, edge label'=$f_R$] (a2) -- [charged scalar] (a3) -- [fermion, edge label'=$F_L$] (a4),
(a2) -- [fermion, quarter left] (d) -- [fermion, quarter left] (a3),
(d) -- [scalar] (e),
};

\end{feynman}
        \end{tikzpicture}
        \caption{Class Two}
        \label{fig:topologyc2}
    \end{subfigure}
    \caption{One-loop topologies for effective Yukawa coupling diagrams before EWSB.}
    \label{fig:topology}
\end{figure}
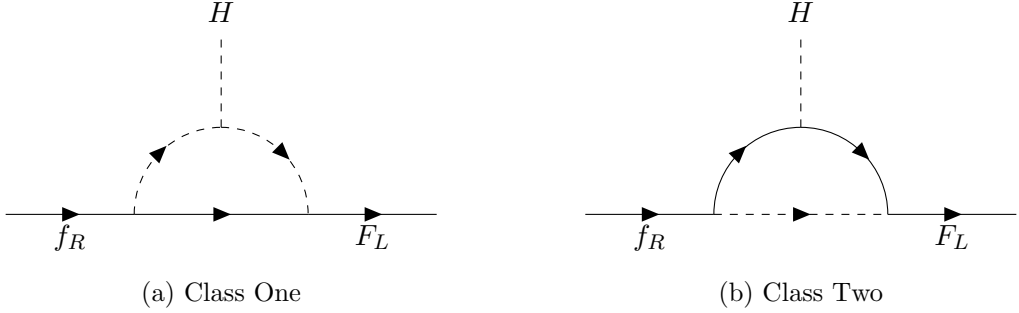

In the framework formulated in~\cite{Baker_2021}, a one-loop mass for a given SM fermion is generated by introducing only new scalars and fermions, and with no modifications to the gauge structure of the theory and only a single Higgs doublet. With these stipulations there are two possible topologies for the one-loop Yukawa diagrams before electroweak symmetry breaking (EWSB), as shown in Figure~\ref{fig:topology}. Class One diagrams feature two exotic scalars which couple to the Higgs, and one exotic fermion, whereas Class Two diagrams feature two exotic fermions which couple to the Higgs, and one exotic scalar.

For a given diagram to exist, the Lagrangian must at a minimum contain the terms corresponding to vertices in the given diagram. Thus
\begin{equation}
    \begin{split}
        -\lagr_{CO} &\supset y_L^F\overline{F_L} \exlab_2^f (\exlab_3^f)_R +  y_R^f \overline{(\exlab_3^f)_L} (\exlab_1^f)^\dagger f_R + aH (\exlab_2^f)^\dagger \exlab_1^f + m_3 \overline{(\exlab_3^f)_L} (\exlab_3^f)_R +\hc , \\
        -\lagr_{CT} &\supset y_L^F \overline{F_L} \exlab_3^f (\exlab_2^f)_R + y_R^f \overline{(\exlab_1^f)_L} (\exlab_3^f)^\dagger f_R + y_H \overline{(\exlab_2^f)_L} H (\exlab_1^f)_R + m_1 \overline{(\exlab_1^f)_L} (\exlab_1^f)_R \\
        & \hspace{8.75cm} + m_2 \overline{(\exlab_2^f)_L} (\exlab_2^f)_R +\hc
    \end{split}
\end{equation}
for Class One and Two diagrams respectively, where we have taken $f_L$ to be the lower component of $F_L$. If it is the upper component, instances of $H$ must be replaced with $\Tilde{H}$. If the exotic has an $L$ or $R$ subscript, then it is a fermion; otherwise it is a scalar.

Lorentz- and gauge-invariance of these couplings restrict the quantum numbers of the exotics to those given in Table~\ref{exotic_gaugenumbers}. Additional symmetries may exist and additional terms may be permitted, though these will be model-dependent. Generally these Lagrangians will admit $U(1)$ symmetries that are broken only softly by the exotic fermion Dirac mass terms or the scalar trilinear term. We demand our models possess at least one such symmetry that forbids the tree level $\overline{F_L} H f_R$ (or $\overline{F_L} \Tilde{H} f_R$) Yukawa term, so that the $f$ mass has its lowest order contribution at one-loop level.\footnote{This is why $\lagr_{CT}$ must not include a $\overline{(\exlab_2^f)_R} H (\exlab_1^f)_L$ term, since no such symmetry could then exist.}

\begin{table}
    \centering
    \begin{tabular}{c|c|cccccc}
Field                 & Spin                  & \multicolumn{4}{c}{$SU(3)_C$}               & $SU(2)_L$            & $U(1)_Y$                 \\
\multicolumn{1}{l|}{} & \multicolumn{1}{l|}{} & LI      & QI        & QII       & QIII      & \multicolumn{1}{l}{} & \multicolumn{1}{l}{}     \\ \hline
$\exlab_1^f$          & $w$                   & $(x,y)$ & $(x,y+1)$ & $(x,y)$   & $(x+1,y)$ & $(z)$                & $Y(f_R) - Y(\exlab_3^f)$ \\
$\exlab_2^f$          & $w$                   & $(x,y)$ & $(x,y+1)$ & $(x,y)$   & $(x+1,y)$ & $(|z\pm 1|)$         & $Y(F_L) - Y(\exlab_3^f)$ \\
$\exlab_3^f$          & $\frac{1}{2} - w$& $(y,x)$ & $(y,x)$   & $(y,x+1)$ & $(y+1,x)$ & $(z)$                & $Y(\exlab_3^f)$         
\end{tabular}
    \caption{Spin and $\mathcal{G}_\mathrm{SM}$ quantum numbers for exotic fields $\exlab^f_{1,2,3}$. Class One diagrams have $w=0$ and Class Two diagrams have $w=\frac{1}{2}$. The non-abelian quantum numbers are given as Dynkin labels, where $x,y,z \in \mathbb{N}$. If $f$ is a lepton then the $SU(3)_C$ assignments will be those given in the LI column. If $f$ is a quark there are instead three possible patterns of assignments, given by QI, QII, and QIII.}
    \label{exotic_gaugenumbers}
\end{table}

\subsection{Minimal models}\label{sec:minimal_models}

We wish to construct models that generate one-loop Yukawas for the $b,c,$ and $\tau$ simultaneously, and to do so with the fewest new degrees of freedom required. In accordance with this aim, the smallest possible number of exotic fields should be introduced. This is accomplished by identifying some exotics in effective Yukawa diagrams for different SM fermions as the same fields (e.g. $X_1^b=X_2^c$). At least three exotics are obviously required, however it is impossible to produce a combined model for $b,c,\tau$ in this framework that introduces only three exotic fields (see Appendix~\ref{sec:three_exotics}). Thus we must instead consider the next-most minimal option, those containing four new fields.

To completely classify this set of models, we first generated the set of all possible identifications that can be made between the $\exlab_i^f$ for $f\in\{b,c,\tau\}$, such that there are four total exotic fields. This is equivalent to listing all possible ways that the new fields may interact with the Higgs and the relevant SM fermions. We then discarded any set of identifications possessing any of the following properties:
\begin{itemize}
    \item Inconsistent spins. For example, requiring $\exlab_1^b = \exlab_1^\tau$ and $\exlab_2^b = \exlab_3^\tau$ simultaneously.
    \item  No possible hypercharge assignments. Identifications between the different $\exlab_i$ produce a set of simultaneous equations for $Y(\exlab_i)$ in accordance with Table~\ref{exotic_gaugenumbers}, for which there sometimes exist no solutions. A simple example of this is requiring $\exlab_1^b = \exlab_1^\tau$ and $\exlab_3^b = \exlab_3^\tau$ simultaneously.
    \item Inconsistent $SU(2)_L$ groupings. For example, requiring $\exlab_1^b = \exlab_2^\tau$ and $\exlab_3^b = \exlab_3^\tau$ simultaneously.
    \item Incapability of forbidding all relevant tree level Yukawa terms. For example, suppose $\exlab_{1}^b = \exlab_2^\tau$ and $\exlab_{2}^b = \exlab_1^\tau$ are both fermions. Then $b_L^\dagger b_R \sim \left(\exlab_2^b \right)_R^\dagger \left( \exlab_1^b \right)_L \sim \left(\exlab_1^\tau \right)_R^\dagger \left(\exlab_2^\tau \right)_L$ under some arbitrary $U(1)$ symmetry. Assuming that $H$ does not transform under this symmetry to avoid complications with the other SM Yukawas, the $\overline{(\exlab_2^{\tau} )_L} H (\exlab_1^{\tau} )_R$  coupling enforces that $(\exlab_2^{\tau} )_L \sim (\exlab_1^{\tau} )_R$ under said symmetry. Thus it would be impossible to forbid the $b$ Yukawa term. 
\end{itemize}

The result of this process is a total of 25 models, which are presented in Table~\ref{tab:minimal_models}. We provide the spin and $\mathcal{G}_\mathrm{SM}$ quantum numbers for each of the four new fields, and their correspondences to fields in the $b, c, \tau$ one-loop Yukawa diagrams. The exotic hypercharge assignments are fixed for all models, though the $SU(3)_C$ and $SU(2)_L$ charges are not. In general there are multiple possible patterns for the $SU(3)_C$ assignments, so for conciseness we give the assignments with the fewest overall degrees of freedom. We provide the possible $SU(2)_L$ assignments in terms of Dynkin labels.

These 25 models may be broadly divided into three separate groups, reflected in our naming convention: pure Class One (CO\#) models, pure Class Two (CT\#) models, and \textit{hybrid} (H\#) models, in which there will exist both Class One and Class Two topology diagrams. Hybrid models are of particular interest since they have not been studied in this framework previously. We analyse one of these, the H2 model, in greater detail in Section~\ref{sec:h2_model} to investigate the generic features these hybrid models may possess.

There are two special ``groups" of models, namely CO1a-d and CO2a-d, that will constitute different models in general, but may produce the same model depending on the chosen $\mathcal{G}_\mathrm{SM}$ charges. The key point is that these contain a scalar with zero hypercharge, which we may denote as $\phi$. If charges are assigned such that $\phi$ is sterile, whenever the Lagrangian contains a term $\phi \mathcal{O}_i$ it must also contain $\phi^\dagger \mathcal{O}_i$, thus the different identifications defining these models would not lead to different structures.

The models CT1, CT3, CO1a-d, CO3, and CO4 are capable of generating a one-loop Dirac Yukawa for a neutrino without any modification to the exotic field content, if a right-handed neutrino is included in the theory. In fact, the model in~\cite{Stockdale:2025sxi} is one such realisation of the CO1a-d models, though it does not explore additional features resulting from the neutrino.

\begin{longtable}[c]{|c|c|c|c|c|c|}
\caption{Minimal models capable of generating one-loop masses and Yukawa couplings for $b,c,\tau$, and their possible quantum number assignments. Hypercharges are fixed for all models, and $SU(2)_L$ assignments are given in terms of Dynkin labels. $SU(3)_C$ assignments are not fixed but the simplest possible assignments are provided, since multiple patterns of Dynkin labels may be possible.\label{tab:minimal_models}}
\\
\hline
Name & Exotics                                                                                                                                                                         & Spin                                                                              & $SU(3)_C$                                                                                                                          & $SU(2)_L$                                                                             & $U(1)_Y$                                                                                                  \\ 
\endfirsthead \hline
Name & Exotics                                                                                                                                                                         & Spin                                                                              & $SU(3)_C$                                                                                                                          & $SU(2)_L$                                                                             & $U(1)_Y$                                                                                                  \\ 
\hline \endhead \hline
H1   & \begin{tabular}[c]{@{}c@{}}$\exlab_1^\tau = \exlab_1^b$\\ $\exlab_1^c$\\ $\exlab_2^\tau = \exlab_2^b = \exlab_3^c$\\ $\exlab_3^\tau = (\exlab_3^b)^c = (\exlab_2^c)^c$\end{tabular}     & \begin{tabular}[c]{@{}c@{}}$0$\\ $\frac{1}{2}$\\ $0$\\ $\frac{1}{2}$\end{tabular} & \begin{tabular}[c]{@{}c@{}}$\overline{\mathbf{3}}$\\ $\overline{\mathbf{3}}$\\ $\overline{\mathbf{3}}$\\ $\mathbf{3}$\end{tabular} & \begin{tabular}[c]{@{}c@{}}$(z)$\\ $(|z \pm 1|)$\\ $(|z \pm 1|)$\\ $(z)$\end{tabular}  & \begin{tabular}[c]{@{}c@{}}$-\frac{2}{3}$\\ $\frac{5}{6}$\\ $-\frac{1}{6}$\\ $-\frac{1}{3}$\end{tabular}  \\ \hline
H2   & \begin{tabular}[c]{@{}c@{}}$\exlab_1^\tau = \exlab_1^b$\\ $\exlab_1^c$\\ $\exlab_2^\tau = \exlab_2^b = \exlab_3^c$\\ $\exlab_3^\tau = (\exlab_3^b)^* = (\exlab_2^c)^*$\end{tabular}     & \begin{tabular}[c]{@{}c@{}}$\frac{1}{2}$\\ $0$\\ $\frac{1}{2}$\\ $0$\end{tabular} & \begin{tabular}[c]{@{}c@{}}$\overline{\mathbf{3}}$\\ $\overline{\mathbf{3}}$\\ $\overline{\mathbf{3}}$\\ $\mathbf{3}$\end{tabular} & \begin{tabular}[c]{@{}c@{}}$(z)$\\ $(|z \pm 1|)$\\ $(|z \pm 1|)$\\ $(z)$\end{tabular}  & \begin{tabular}[c]{@{}c@{}}$-\frac{2}{3}$\\ $\frac{5}{6}$\\ $-\frac{1}{6}$\\ $-\frac{1}{3}$\end{tabular}  \\ \hline
H3   & \begin{tabular}[c]{@{}c@{}}$\exlab_1^\tau = (\exlab_2^b)^* = (\exlab_3^c)^*$\\ $\exlab_2^\tau = (\exlab_1^b)^*$\\ $\exlab_3^\tau = \exlab_1^c$\\ $\exlab_3^b = \exlab_2^c$\end{tabular} & \begin{tabular}[c]{@{}c@{}}$0$\\ $0$\\ $\frac{1}{2}$\\ $\frac{1}{2}$\end{tabular} & \begin{tabular}[c]{@{}c@{}}$\mathbf{3}$\\ $\mathbf{3}$\\ $\overline{\mathbf{3}}$\\ $\overline{\mathbf{3}}$\end{tabular}            & \begin{tabular}[c]{@{}c@{}}$(z)$\\ $(|z \pm 1|)$\\ $(z)$\\ $(|z \pm 1|)$\end{tabular} & \begin{tabular}[c]{@{}c@{}}$-\frac{5}{6}$\\ $-\frac{1}{3}$\\ $-\frac{1}{6}$\\ $-\frac{2}{3}$\end{tabular} \\ \hline
H4   & \begin{tabular}[c]{@{}c@{}}$\exlab_1^\tau = (\exlab_2^b)^c = (\exlab_3^c)^c$\\ $\exlab_2^\tau = (\exlab_1^b)^*$\\ $\exlab_3^\tau = \exlab_1^c$\\ $\exlab_3^b = \exlab_2^c$\end{tabular} & \begin{tabular}[c]{@{}c@{}}$\frac{1}{2}$\\ $\frac{1}{2}$\\ $0$\\ $0$\end{tabular} & \begin{tabular}[c]{@{}c@{}}$\mathbf{3}$\\ $\mathbf{3}$\\ $\overline{\mathbf{3}}$\\ $\overline{\mathbf{3}}$\end{tabular}            & \begin{tabular}[c]{@{}c@{}}$(z)$\\ $(|z \pm 1|)$\\ $(z)$\\ $(|z \pm 1|)$\end{tabular} & \begin{tabular}[c]{@{}c@{}}$-\frac{5}{6}$\\ $-\frac{1}{3}$\\ $-\frac{1}{6}$\\ $-\frac{2}{3}$\end{tabular} \\ \hline
H5   & \begin{tabular}[c]{@{}c@{}}$\exlab_1^\tau = (\exlab_1^c)^*$\\ $\exlab_1^b$\\ $\exlab_2^\tau = (\exlab_3^b)^* = (\exlab_2^c)^*$\\ $\exlab_3^\tau = \exlab_2^b = \exlab_3^c$\end{tabular} & \begin{tabular}[c]{@{}c@{}}$0$\\ $\frac{1}{2}$\\ $0$\\ $\frac{1}{2}$\end{tabular} & \begin{tabular}[c]{@{}c@{}}$\mathbf{3}$\\ $\overline{\mathbf{3}}$\\ $\mathbf{3}$\\ $\overline{\mathbf{3}}$\end{tabular}            & \begin{tabular}[c]{@{}c@{}}$(z)$\\ $(|z \pm 1|)$\\ $(|z \pm 1|)$\\ $(z)$\end{tabular} & \begin{tabular}[c]{@{}c@{}}$-\frac{5}{6}$\\ $-\frac{2}{3}$\\ $-\frac{1}{3}$\\ $-\frac{1}{6}$\end{tabular} \\ \hline
H6   & \begin{tabular}[c]{@{}c@{}}$\exlab_1^\tau = \exlab_3^b = \exlab_2^c$\\ $\exlab_2^\tau = \exlab_1^c$\\ $\exlab_2^b = \exlab_3^c$\\ $\exlab_3^\tau = (\exlab_1^b)^c$\end{tabular}         & \begin{tabular}[c]{@{}c@{}}$0$\\ $0$\\ $\frac{1}{2}$\\ $\frac{1}{2}$\end{tabular} & \begin{tabular}[c]{@{}c@{}}$\overline{\mathbf{3}}$\\ $\overline{\mathbf{3}}$\\ $\overline{\mathbf{3}}$\\ $\mathbf{3}$\end{tabular} & \begin{tabular}[c]{@{}c@{}}$(z)$\\ $(|z \pm 1|)$\\ $(|z \pm 1|)$\\ $(z)$\end{tabular} & \begin{tabular}[c]{@{}c@{}}$-\frac{2}{3}$\\ $-\frac{1}{6}$\\ $\frac{5}{6}$\\ $-\frac{1}{3}$\end{tabular}  \\ \hline
H7   & \begin{tabular}[c]{@{}c@{}}$\exlab_1^\tau = (\exlab_3^c)^c$\\ $\exlab_1^b = \exlab_2^c$\\ $\exlab_2^\tau = (\exlab_3^b)^c$\\ $\exlab_3^\tau = \exlab_2^b = \exlab_1^c$\end{tabular}     & \begin{tabular}[c]{@{}c@{}}$\frac{1}{2}$\\ $0$\\ $\frac{1}{2}$\\ $0$\end{tabular} & \begin{tabular}[c]{@{}c@{}}$\mathbf{3}$\\ $\overline{\mathbf{3}}$\\ $\mathbf{3}$\\ $\overline{\mathbf{3}}$\end{tabular}            & \begin{tabular}[c]{@{}c@{}}$(z)$\\ $(|z \pm 1|)$\\ $(|z \pm 1|)$\\ $(z)$\end{tabular} & \begin{tabular}[c]{@{}c@{}}$-\frac{5}{6}$\\ $-\frac{2}{3}$\\ $-\frac{1}{3}$\\ $-\frac{1}{6}$\end{tabular} \\ \hline
H8   & \begin{tabular}[c]{@{}c@{}}$\exlab_1^\tau = \exlab_3^b$\\ $\exlab_2^b = \exlab_1^c$\\ $\exlab_2^\tau = \exlab_3^c$\\ $\exlab_3^\tau = (\exlab_1^b)^c = (\exlab_2^c)^c$\end{tabular}     & \begin{tabular}[c]{@{}c@{}}$0$\\ $\frac{1}{2}$\\ $0$\\ $\frac{1}{2}$\end{tabular} & \begin{tabular}[c]{@{}c@{}}$\overline{\mathbf{3}}$\\ $\overline{\mathbf{3}}$\\ $\overline{\mathbf{3}}$\\ $\mathbf{3}$\end{tabular} & \begin{tabular}[c]{@{}c@{}}$(z)$\\ $(|z \pm 1|)$\\ $(|z \pm 1|)$\\ $(z)$\end{tabular}   & \begin{tabular}[c]{@{}c@{}}$-\frac{2}{3}$\\ $\frac{5}{6}$\\ $-\frac{1}{6}$\\ $-\frac{1}{3}$\end{tabular}  \\ \hline
CO1a    & \begin{tabular}[c]{@{}c@{}}$\exlab_1^\tau = \exlab_1^b = (\exlab_1^c)^*$\\ $\exlab_2^\tau = \exlab_2^b = \exlab_2^c$\\ $\exlab_3^\tau$\\ $\exlab_3^b = \exlab_3^c$\end{tabular}             & \begin{tabular}[c]{@{}c@{}}$0$\\ $0$\\ $\frac{1}{2}$\\ $\frac{1}{2}$\end{tabular} & \begin{tabular}[c]{@{}c@{}}$\mathbf{1}$\\ $\mathbf{1}$\\ $\mathbf{1}$\\ $\mathbf{3}$\end{tabular}                                  & \begin{tabular}[c]{@{}c@{}}$(z)$\\ $(|z \pm 1|)$\\ $(z)$\\ $(z)$\end{tabular}          & \begin{tabular}[c]{@{}c@{}}$-\frac{1}{2}$\\ $0$\\ $-\frac{1}{2}$\\ $\frac{1}{6}$\end{tabular}             \\ \hline
CO1b    & \begin{tabular}[c]{@{}c@{}}$\exlab_1^\tau = \exlab_1^b = (\exlab_1^c)^*$\\ $\exlab_2^\tau = (\exlab_2^b)^* = \exlab_2^c$\\ $\exlab_3^\tau$\\ $\exlab_3^b = \exlab_3^c$\end{tabular}         & \begin{tabular}[c]{@{}c@{}}$0$\\ $0$\\ $\frac{1}{2}$\\ $\frac{1}{2}$\end{tabular} & \begin{tabular}[c]{@{}c@{}}$\mathbf{1}$\\ $\mathbf{1}$\\ $\mathbf{1}$\\ $\mathbf{3}$\end{tabular}                                  & \begin{tabular}[c]{@{}c@{}}$(z)$\\ $(|z \pm 1|)$\\ $(z)$\\ $(z)$\end{tabular}          & \begin{tabular}[c]{@{}c@{}}$-\frac{1}{2}$\\ $0$\\ $-\frac{1}{2}$\\ $\frac{1}{6}$\end{tabular}             \\ \hline
CO1c    & \begin{tabular}[c]{@{}c@{}}$\exlab_1^\tau = \exlab_1^b = (\exlab_1^c)^*$\\ $\exlab_2^\tau = \exlab_2^b = (\exlab_2^c)^*$\\ $\exlab_3^\tau$\\ $\exlab_3^b = \exlab_3^c$\end{tabular}         & \begin{tabular}[c]{@{}c@{}}$0$\\ $0$\\ $\frac{1}{2}$\\ $\frac{1}{2}$\end{tabular} & \begin{tabular}[c]{@{}c@{}}$\mathbf{1}$\\ $\mathbf{1}$\\ $\mathbf{1}$\\ $\mathbf{3}$\end{tabular}                                  & \begin{tabular}[c]{@{}c@{}}$(z)$\\ $(|z \pm 1|)$\\ $(z)$\\ $(z)$\end{tabular}          & \begin{tabular}[c]{@{}c@{}}$-\frac{1}{2}$\\ $0$\\ $-\frac{1}{2}$\\ $\frac{1}{6}$\end{tabular}             \\ \hline
CO1d    & \begin{tabular}[c]{@{}c@{}}$\exlab_1^\tau = \exlab_1^b = (\exlab_1^c)^*$\\ $\exlab_2^\tau = (\exlab_2^b)^* = (\exlab_2^c)^*$\\ $\exlab_3^\tau$\\ $\exlab_3^b = \exlab_3^c$\end{tabular}     & \begin{tabular}[c]{@{}c@{}}$0$\\ $0$\\ $\frac{1}{2}$\\ $\frac{1}{2}$\end{tabular} & \begin{tabular}[c]{@{}c@{}}$\mathbf{1}$\\ $\mathbf{1}$\\ $\mathbf{1}$\\ $\mathbf{3}$\end{tabular}                                  & \begin{tabular}[c]{@{}c@{}}$(z)$\\ $(|z \pm 1|)$\\ $(z)$\\ $(z)$\end{tabular}         & \begin{tabular}[c]{@{}c@{}}$-\frac{1}{2}$\\ $0$\\ $-\frac{1}{2}$\\ $\frac{1}{6}$\end{tabular}             \\ \hline
CO2a    & \begin{tabular}[c]{@{}c@{}}$\exlab_1^\tau = (\exlab_2^b)^* = (\exlab_2^c)^*$\\ $\exlab_2^\tau = (\exlab_1^b)^* = \exlab_1^c$\\ $\exlab_3^\tau$\\ $\exlab_3^b = \exlab_3^c$\end{tabular}     & \begin{tabular}[c]{@{}c@{}}$0$\\ $0$\\ $\frac{1}{2}$\\ $\frac{1}{2}$\end{tabular} & \begin{tabular}[c]{@{}c@{}}$\mathbf{1}$\\ $\mathbf{1}$\\ $\mathbf{1}$\\ $\mathbf{3}$\end{tabular}                                  & \begin{tabular}[c]{@{}c@{}}$(z)$\\ $(|z \pm 1|)$\\ $(z)$\\ $(|z \pm 1|)$\end{tabular} & \begin{tabular}[c]{@{}c@{}}$0$\\ $\frac{1}{2}$\\ $-1$\\ $\frac{1}{6}$\end{tabular}                        \\ \hline
CO2b    & \begin{tabular}[c]{@{}c@{}}$\exlab_1^\tau = \exlab_2^b = (\exlab_2^c)^*$\\ $\exlab_2^\tau = (\exlab_1^b)^* = \exlab_1^c$\\ $\exlab_3^\tau$\\ $\exlab_3^b = \exlab_3^c$\end{tabular}         & \begin{tabular}[c]{@{}c@{}}$0$\\ $0$\\ $\frac{1}{2}$\\ $\frac{1}{2}$\end{tabular} & \begin{tabular}[c]{@{}c@{}}$\mathbf{1}$\\ $\mathbf{1}$\\ $\mathbf{1}$\\ $\mathbf{3}$\end{tabular}                                  & \begin{tabular}[c]{@{}c@{}}$(z)$\\ $(|z \pm 1|)$\\ $(z)$\\ $(|z \pm 1|)$\end{tabular} & \begin{tabular}[c]{@{}c@{}}$0$\\ $\frac{1}{2}$\\ $-1$\\ $\frac{1}{6}$\end{tabular}                        \\ \hline
CO2c    & \begin{tabular}[c]{@{}c@{}}$\exlab_1^\tau = (\exlab_2^b)^* = \exlab_2^c$\\ $\exlab_2^\tau = (\exlab_1^b)^* = \exlab_1^c$\\ $\exlab_3^\tau$\\ $\exlab_3^b = \exlab_3^c$\end{tabular}         & \begin{tabular}[c]{@{}c@{}}$0$\\ $0$\\ $\frac{1}{2}$\\ $\frac{1}{2}$\end{tabular} & \begin{tabular}[c]{@{}c@{}}$\mathbf{1}$\\ $\mathbf{1}$\\ $\mathbf{1}$\\ $\mathbf{3}$\end{tabular}                                  & \begin{tabular}[c]{@{}c@{}}$(z)$\\ $(|z \pm 1|)$\\ $(z)$\\ $(|z \pm 1|)$\end{tabular} & \begin{tabular}[c]{@{}c@{}}$0$\\ $\frac{1}{2}$\\ $-1$\\ $\frac{1}{6}$\end{tabular}                        \\ \hline
CO2d    & \begin{tabular}[c]{@{}c@{}}$\exlab_1^\tau = \exlab_2^b = \exlab_2^c$\\ $\exlab_2^\tau = (\exlab_1^b)^* = \exlab_1^c$\\ $\exlab_3^\tau$\\ $\exlab_3^b = \exlab_3^c$\end{tabular}             & \begin{tabular}[c]{@{}c@{}}$0$\\ $0$\\ $\frac{1}{2}$\\ $\frac{1}{2}$\end{tabular} & \begin{tabular}[c]{@{}c@{}}$\mathbf{1}$\\ $\mathbf{1}$\\ $\mathbf{1}$\\ $\mathbf{3}$\end{tabular}                                  & \begin{tabular}[c]{@{}c@{}}$(z)$\\ $(|z \pm 1|)$\\ $(z)$\\ $(|z \pm 1|)$\end{tabular} & \begin{tabular}[c]{@{}c@{}}$0$\\ $\frac{1}{2}$\\ $-1$\\ $\frac{1}{6}$\end{tabular}                        \\ \hline
CO3     & \begin{tabular}[c]{@{}c@{}}$\exlab_1^\tau = \exlab_1^b$\\ $\exlab_1^c$\\ $\exlab_2^\tau = \exlab_2^b = \exlab_2^c$\\ $\exlab_3^\tau = (\exlab_3^b)^c = (\exlab_3^c)^c$\end{tabular}         & \begin{tabular}[c]{@{}c@{}}$0$\\ $0$\\ $0$\\ $\frac{1}{2}$\end{tabular}           & \begin{tabular}[c]{@{}c@{}}$\overline{\mathbf{3}}$\\ $\overline{\mathbf{3}}$\\ $\overline{\mathbf{3}}$\\ $\mathbf{3}$\end{tabular} & \begin{tabular}[c]{@{}c@{}}$(z)$\\ $(z)$\\ $(|z \pm 1|)$\\ $(z)$\end{tabular}         & \begin{tabular}[c]{@{}c@{}}$-\frac{2}{3}$\\ $\frac{1}{3}$\\ $-\frac{1}{6}$\\ $- \frac{1}{3}$\end{tabular} \\ \hline
CO4  & \begin{tabular}[c]{@{}c@{}}$\exlab_1^\tau = (\exlab_1^c)^*$\\ $\exlab_1^b$\\ $\exlab_2^\tau = (\exlab_2^b)^* = (\exlab_2^c)^*$\\ $\exlab_3^\tau = \exlab_3^b = \exlab_3^c$\end{tabular}     & \begin{tabular}[c]{@{}c@{}}$0$\\ $0$\\ $0$\\ $\frac{1}{2}$\end{tabular}                     & \begin{tabular}[c]{@{}c@{}}$\mathbf{3}$\\ $\overline{\mathbf{3}}$\\ $\mathbf{3}$\\ $\overline{\mathbf{3}}$\end{tabular}            & \begin{tabular}[c]{@{}c@{}}$(z)$\\ $(z)$\\ $(| z \pm 1|)$\\ $(z)$\end{tabular}        & \begin{tabular}[c]{@{}c@{}}$-\frac{5}{6}$\\ $-\frac{1}{6}$\\ $-\frac{1}{3}$\\ $-\frac{1}{6}$\end{tabular} \\ \hline
CO5  & \begin{tabular}[c]{@{}c@{}}$\exlab_1^\tau = \exlab_1^b = (\exlab_1^c)^*$\\ $\exlab_2^\tau = \exlab_2^b = (\exlab_2^c)^*$\\ $\exlab_3^\tau = (\exlab_3^b)^c$\\ $\exlab_3^c$\end{tabular}     & \begin{tabular}[c]{@{}c@{}}$0$\\ $0$\\ $\frac{1}{2}$\\ $\frac{1}{2}$\end{tabular}           & \begin{tabular}[c]{@{}c@{}}$\overline{\mathbf{3}}$\\ $\overline{\mathbf{3}}$\\ $\mathbf{3}$\\ $\mathbf{1}$\end{tabular}            & \begin{tabular}[c]{@{}c@{}}$(z)$\\ $(|z \pm 1|)$\\ $(z)$\\ $(z)$\end{tabular}         & \begin{tabular}[c]{@{}c@{}}$-\frac{2}{3}$\\ $-\frac{1}{6}$\\ $-\frac{1}{3}$\\ $0$\end{tabular}            \\ \hline
CO6  & \begin{tabular}[c]{@{}c@{}}$\exlab_1^\tau = \exlab_1^b = (\exlab_1^c)^*$\\ $\exlab_2^\tau = \exlab_2^b = (\exlab_2^c)^*$\\ $\exlab_3^\tau = \exlab_3^c$\\ $\exlab_3^b$\end{tabular}         & \begin{tabular}[c]{@{}c@{}}$0$\\ $0$\\ $\frac{1}{2}$\\ $\frac{1}{2}$\end{tabular}           & \begin{tabular}[c]{@{}c@{}}$\mathbf{3}$\\ $\mathbf{3}$\\ $\overline{\mathbf{3}}$\\ $\mathbf{1}$\end{tabular}                       & \begin{tabular}[c]{@{}c@{}}$(z)$\\ $(|z \pm 1|)$\\ $(z)$\\ $(z)$\end{tabular}         & \begin{tabular}[c]{@{}c@{}}$-\frac{5}{6}$\\ $-\frac{1}{3}$\\ $-\frac{1}{6}$\\ $\frac{1}{2}$\end{tabular}  \\ \hline
CO7  & \begin{tabular}[c]{@{}c@{}}$\exlab_1^\tau = \exlab_1^b = \exlab_2^c$\\ $\exlab_2^\tau = \exlab_2^b = \exlab_1^c$\\ $\exlab_3^\tau = (\exlab_3^b)^c$\\ $\exlab_3^c$\end{tabular}             & \begin{tabular}[c]{@{}c@{}}$0$\\ $0$\\ $\frac{1}{2}$\\ $\frac{1}{2}$\end{tabular}           & \begin{tabular}[c]{@{}c@{}}$\overline{\mathbf{3}}$\\ $\overline{\mathbf{3}}$\\ $\mathbf{3}$\\ $\overline{\mathbf{3}}$\end{tabular} & \begin{tabular}[c]{@{}c@{}}$(z)$\\ $(|z \pm 1|)$\\ $(z)$\\ $(|z \pm 1|)$\end{tabular} & \begin{tabular}[c]{@{}c@{}}$-\frac{2}{3}$\\ $-\frac{1}{6}$\\ $-\frac{1}{3}$\\ $\frac{5}{6}$\end{tabular}  \\ \hline
CO8  & \begin{tabular}[c]{@{}c@{}}$\exlab_1^\tau = (\exlab_2^b)^* = (\exlab_1^c)^*$\\ $\exlab_2^\tau = (\exlab_1^b)^* = (\exlab_2^c)^*$\\ $\exlab_3^\tau = \exlab_3^c$\\ $\exlab_3^b$\end{tabular} & \begin{tabular}[c]{@{}c@{}}$0$\\ $0$\\ $\frac{1}{2}$\\ $\frac{1}{2}$\end{tabular}           & \begin{tabular}[c]{@{}c@{}}$\mathbf{3}$\\ $\mathbf{3}$\\ $\overline{\mathbf{3}}$\\ $\overline{\mathbf{3}}$\end{tabular}            & \begin{tabular}[c]{@{}c@{}}$(z)$\\ $(|z \pm 1|)$\\ $(z)$\\ $(|z \pm 1|)$\end{tabular} & \begin{tabular}[c]{@{}c@{}}$-\frac{5}{6}$\\ $-\frac{1}{3}$\\ $-\frac{1}{6}$\\ $-\frac{2}{3}$\end{tabular} \\ \hline
CT1  & \begin{tabular}[c]{@{}c@{}}$\exlab_1^\tau = \exlab_1^b = (\exlab_1^c)^c$\\ $\exlab_2^\tau = \exlab_2^b = \exlab_2^c$\\ $\exlab_3^\tau$\\ $\exlab_3^b = \exlab_3^c$\end{tabular}             & \begin{tabular}[c]{@{}c@{}}$\frac{1}{2}$\\ $\frac{1}{2}$\\ $0$\\ $0$\end{tabular}           & \begin{tabular}[c]{@{}c@{}}$\mathbf{1}$\\ $\mathbf{1}$\\ $\mathbf{1}$\\ $\mathbf{3}$\end{tabular}                                  & \begin{tabular}[c]{@{}c@{}}$(z)$\\ $(|z \pm 1|)$\\ $(z)$\\ $(z)$\end{tabular}          & \begin{tabular}[c]{@{}c@{}}$-\frac{1}{2}$\\ $0$\\ $-\frac{1}{2}$\\ $\frac{1}{6}$\end{tabular}             \\ \hline
CT2  & \begin{tabular}[c]{@{}c@{}}$\exlab_1^\tau = (\exlab_2^b)^c = (\exlab_2^c)^c$\\ $\exlab_2^\tau = (\exlab_1^b)^c = \exlab_1^c$\\ $\exlab_3^\tau$\\ $\exlab_3^b = \exlab_3^c$\end{tabular}     & \begin{tabular}[c]{@{}c@{}}$\frac{1}{2}$\\ $\frac{1}{2}$\\ $0$\\ $0$\end{tabular}           & \begin{tabular}[c]{@{}c@{}}$\mathbf{1}$\\ $\mathbf{1}$\\ $\mathbf{1}$\\ $\mathbf{3}$\end{tabular}                                  & \begin{tabular}[c]{@{}c@{}}$(z)$\\ $(|z \pm 1|)$\\ $(z)$\\ $(|z \pm 1|)$\end{tabular} & \begin{tabular}[c]{@{}c@{}}$0$\\ $\frac{1}{2}$\\ $-1$\\ $\frac{1}{6}$\end{tabular}                        \\ \hline
CT3  & \begin{tabular}[c]{@{}c@{}}$\exlab_1^\tau = \exlab_1^b$\\ $\exlab_1^c$\\ $\exlab_2^\tau = \exlab_2^b = \exlab_2^c$\\ $\exlab_3^\tau = (\exlab_3^b)^* = (\exlab_3^c)^*$\end{tabular}         & \begin{tabular}[c]{@{}c@{}}$\frac{1}{2}$\\ $\frac{1}{2}$\\ $\frac{1}{2}$\\ $0$\end{tabular} & \begin{tabular}[c]{@{}c@{}}$\overline{\mathbf{3}}$\\ $\overline{\mathbf{3}}$\\ $\overline{\mathbf{3}}$\\ $\mathbf{3}$\end{tabular} & \begin{tabular}[c]{@{}c@{}}$(z)$\\ $(z)$\\ $(|z \pm 1|)$\\ $(z)$\end{tabular}         & \begin{tabular}[c]{@{}c@{}}$-\frac{2}{3}$\\ $\frac{1}{3}$\\ $-\frac{1}{6}$\\ $- \frac{1}{3}$\end{tabular} \\ \hline
\end{longtable}

\section{The H2 model}\label{sec:h2_model}

To see in detail how one of these models works, we choose to examine the H2 case. This provides a contrasting benchmark model to the one we analysed in Ref.~\cite{Stockdale:2025sxi}.

\subsection{Model structure}

To simplify the notation, we denote $\chi = \exlab_1^\tau = \exlab_1^b$, $\psi = \exlab_2^\tau = \exlab_2^b=\exlab_3^c$ as the vectorlike fermions and $\eta = \exlab_1^c$, $\phi = (\exlab_3^\tau)^* = \exlab_3^b = \exlab_2^c$ as the complex scalars in this model. The defining set of identifications requires our Lagrangian to contain
\begin{equation}
\begin{split}
    - \lagr \supset y_L^L \overline{L_L} \phi^\dagger \psi_R + y_L^{Q_i} \overline{Q^i_L} \phi \psi_R + y_R^\tau \overline{\chi_L} \phi \tau_R + y_R^b \overline{\chi_L} \phi^\dagger b_R + y_R^c \overline{\psi_L} \eta^\dagger c_R& \\
    + y_H \overline{\psi_L} H \chi_R + a H \eta^\dagger \phi + m_\psi \overline{\psi_L} \psi_R + m_\chi \overline{\chi_L} \chi_R + \hc&
\end{split}
\end{equation}
The quantum numbers of the relevant fields are given in Table~\ref{tab:h2qn}, choosing the simplest possible set of assignments. This Lagrangian possesses three accidental symmetries: extended lepton and baryon numbers, and exotic parity $\mathcal{Z}_2$. It also admits three symmetries that are broken only softly: $U(1)_\psi$, $U(1)_\chi$, and $U(1)_a$. These are highly important because $U(1)_\psi$ forbids tree-level Yukawa coupling terms for $b$, $c$ and $\tau$, thus banishing the generation of tree-level masses for all three relevant species, while $U(1)_\chi$ forbids $b$ and  $\tau$ tree-level masses, and $U(1)_a$ forbids a tree-level $c$ mass. We choose to impose all three on the dimension-four terms in the theory, but permit the soft-breaking terms that are required for loop-level mass generation. We do not consider any additional terms coupling the exotics to the lighter fermions, so as to remain agnostic to the physics of a complete model treating all three generations.

To avoid bounds on coloured relics, one may include $\mathcal{Z}_2$ violating terms in the Lagrangian, which will necessarily break the extended lepton and baryon number symmetries to the combination $B-L$. These $\mathcal{Z}_2$ violating terms, such as $\overline{\psi_L} (Q_L)^c$ and $\overline{\chi_R} (c_R)^c$, will induce mixing between the exotic and SM fermions. This will not reintroduce tree level masses corresponding to $m_b$ or $m_c$, though it will extend the number of parameters in the theory and alter the mixing behaviour in a way highly specific to this particular model. It would also lead to a violation of unitarity in the CKM sub-matrix, and to proton decay. Alternatively, if the $\mathcal{Z}_2$ remains unbroken, the stable coloured relics may be inflated away, given that we expect the exotic masses to be $\gtrsim 1$ TeV. In what follows we only consider mixing between the exotics, since this generalises more simply across the catalogue, corresponding either to an unbroken $\mathcal{Z}_2$ or small $\mathcal{Z}_2$ violating terms. 

In the absence of $\mathcal{Z}_2$ violating terms, the only additional terms that must be included are those in the scalar potential, which takes the form
\begin{equation}
    \begin{split}
        V =& V_\mathrm{SM} + \mu_\phi^2 \phi^\dagger \phi + \frac{\lambda_\phi}{2} (\phi^\dagger \phi)^2 + \mu_\eta^2 \eta^\dagger \eta + \frac{\lambda_\eta}{2} (\eta^\dagger \eta)^2  + \lambda_1(\phi^\dagger \phi) (H^\dagger H) \\
        &+ \lambda_2(\eta^\dagger \eta)(H^\dagger H)+ \lambda_3(\phi^\dagger \phi)(\eta^\dagger \eta) + \lambda_4(\phi^\dagger \eta)(\eta^\dagger \phi)  + [aH \eta^\dagger \phi + \hc]
    \end{split}
\end{equation}
After EWSB, the scalar masses will be given by
\begin{equation}
\begin{split}
    m_\phi^2 = \mu_\phi^2 + \frac{\lambda_1 v^2}{2} \\
    m_\eta^2 = \mu_\eta^2 + \frac{\lambda_2 v^2}{2}
\end{split}
\end{equation}
The values of other new parameters in the potential, excluding $a$, have no impact on the constraints we consider (at lowest order). We assume that neither $\phi$ nor $\eta$ obtain a VEV.

\begin{table}[t]
\centering
\begin{tabular}{c|cccccccccccc}
                & $\tau_R$     & $L_L$          & $b_R$          & $c_R$         & $Q_L^{2,3}$   & $H$           & $\psi_L$                & $\psi_R$               & $\chi_L$                & $\chi_R$                & $\phi$                  & $\eta$                  \\ \hline
$SU(3)_C$       & $\mathbf{1}$ & $\mathbf{1}$   & $\mathbf{3}$   & $\mathbf{3}$  & $\mathbf{3}$  & $\mathbf{1}$  & $\overline{\mathbf{3}}$ & $\overline{\mathbf{3}}$ & $\overline{\mathbf{3}}$ & $\overline{\mathbf{3}}$ & $\overline{\mathbf{3}}$ & $\overline{\mathbf{3}}$ \\
$SU(2)_L$       & $\mathbf{1}$ & $\mathbf{2}$   & $\mathbf{1}$   & $\mathbf{1}$  & $\mathbf{2}$  & $\mathbf{2}$  & $\mathbf{2}$            & $\mathbf{2}$            & $\mathbf{1}$            & $\mathbf{1}$            & $\mathbf{1}$            & $\mathbf{2}$            \\
$U(1)_Y$        & $-1$         & $-\frac{1}{2}$ & $-\frac{1}{3}$ & $\frac{2}{3}$ & $\frac{1}{6}$ & $\frac{1}{2}$ & $-\frac{1}{6}$          & $-\frac{1}{6}$          & $-\frac{2}{3}$          & $-\frac{2}{3}$          & $\frac{1}{3}$           & $\frac{5}{6}$           \\ \hline
$L$             & $1$          & $1$            & $0$            & $0$           & $0$           & $0$           & $\frac{1}{2}$           & $\frac{1}{2}$           & $\frac{1}{2}$           & $\frac{1}{2}$           & $-\frac{1}{2}$          & $-\frac{1}{2}$          \\
$3B$            & $0$          & $0$            & $1$            & $1$           & $1$           & $0$           & $\frac{1}{2}$           & $\frac{1}{2}$           & $\frac{1}{2}$           & $\frac{1}{2}$           & $\frac{1}{2}$           & $\frac{1}{2}$           \\
$\mathcal{Z}_2$ & $1$          & $1$            & $1$            & $1$           & $1$           & $1$           & $-1$                    & $-1$                    & $-1$                    & $-1$                    & $-1$                    & $-1$                    \\ \hline
$U(1)_\psi$     & $1$          & $0$            & $1$            & $1$           & $0$           & $0$           & $1$                     & $0$                     & $1$                     & $1$                     & $0$                     & $0$                     \\
$U(1)_\chi$     & $1$          & $0$            & $1$            & $0$           & $0$           & $0$           & $0$                     & $0$                     & $1$                     & $0$                     & $0$                     & $0$                     \\
$U(1)_a$        & $0$          & $0$            & $0$            & $1$           & $0$           & $0$           & $0$                     & $0$                     & $0$                     & $0$                     & $0$                     & $1$                    
\end{tabular}
\caption{Quantum numbers for the exotics and the relevant SM fields in the H2 model. }
\label{tab:h2qn}
\end{table}

\subsection{Radiative masses and Yukawa couplings}

The exotic mixing behaviour is the same as in the standard single class models in~\cite{Baker_2021}. After EWSB, the trilinear coupling will cause $\phi$ to mix with the lower component of $\eta$ with angle $\theta_s$, and the $y_H$ term will cause $\chi$ to mix with the lower component of $\psi$ with angles $\theta_L$ and $\theta_R$. The relevant mass terms may be written
\begin{equation}
    -\lagr \supset \mqty(\phi^\dagger & (\eta^{low})^\dagger) \mqty(m_\phi^2 & \frac{av}{\sqrt{2}} \\ \frac{av}{\sqrt{2}} & m_\eta^2) \mqty(\phi \\ \eta^{low}) + \left( \mqty(\overline{\psi_L^{low}} & \overline{\chi_L}) \mqty(m_\psi & \frac{y_H v}{\sqrt{2}} \\ 0 & m_\chi) \mqty(\psi_R^{low} \\ \chi_R) +\hc \right)
\end{equation}
Defining the mass eigenstates as
\begin{equation}
    \begin{split}
        \mqty(\varphi_1 \\ \varphi_2) &= \mqty(\cos \theta_s & -\sin \theta_s \\ \sin \theta_s & \cos \theta_s) \mqty(\phi \\ \eta^{low})~, \\
        \mqty(\zeta_1 \\ \zeta_2)_{L,R} &= \mqty(\cos \theta_{L,R} & -\sin \theta_{L,R} \\ \sin \theta_{L,R} & \cos \theta_{L,R}) \mqty(\psi^{low} \\ \chi)_{L,R}~,
    \end{split}
\end{equation}
the resultant mass eigenvalues are
\begin{equation}
    \begin{split}
        m_{\varphi_{1,2}}^2 &= \frac{1}{2} \left( m_\phi^2 + m_\eta^2 \mp \sqrt{\left(m_\eta^2 - m_\phi^2  \right)^2 + 2 a^2v^2} \right)~,\\
        m_{\zeta_{1,2}}^2 &= \frac{1}{4} \left( 2m_\psi^2 + 2m_\chi^2 + y_H^2 v^2 \mp \sqrt{4\left(m_\psi^2 - m_\chi^2  \right)^2 + y_H^2 v^2 (4m_\psi^2 + 4m_\chi^2 + y_H^2 v^2)} \right)~.
    \end{split}
\end{equation}
We also obtain the relationships
\begin{equation}
    \begin{split}
        \sin 2 \theta_s &= \frac{\sqrt{2} a v}{m_{\varphi_2}^2 - m_{\varphi_1}^2} \, , \\
        \tan 2\theta_{L,R} &=  \frac{y_H v}{\sqrt{2}}\frac{4 m_{\chi,\psi}}{2(m_\chi^2 - m_\psi^2) \mp y_H^2 v^2} \, , \\
        \cos \theta_{L,R} \sin \theta_{R,L} &= \frac{y_H v}{\sqrt{2}} \frac{m_{\zeta_{1_,2}}}{m_{\zeta_2}^2 - m_{\zeta_1}^2} \, .
    \end{split}
\end{equation}

\begin{figure}[t]
    \centering
    \begin{tikzpicture}
        \begin{feynman}
\vertex (a1);
\vertex[right=1.3cm of a1] (a2);
\vertex[right=2.2cm of a2] (a3);
\vertex[right=1.3cm of a3] (a4);

\diagram* {
(a1) -- [fermion, edge label'=$f_R$] (a2) -- [with arrow=0.5, half right, edge label'=$\zeta_{1,2}$] (a3) -- [fermion, edge label'=$f_L$] (a4),
(a2) -- [scalar, half left, edge label=$\varphi_{1,2}$] (a3),
};

\end{feynman}
        \end{tikzpicture}
    \caption{Mass diagrams at one-loop for $b$, $c$, and $\tau$ in the H2 model after EWSB. Each mass will have contributions from these diagrams that differ only by vertex and group theory factors.}
    \label{fig:loop_masses}
\end{figure}
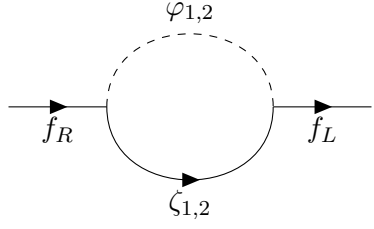

The one-loop mass diagrams for $b,c,\tau$ are as shown in Figure~\ref{fig:loop_masses}, and contributions from each diagram will differ only by various vertex factors. Specifically, the masses at one-loop are
\begin{equation}\label{eq:SMf_masses}
    \begin{split}
        m_c &= \frac{2y_L^{Q_2} y_R^c}{16\pi^2} \frac{v}{\sqrt{2}} \frac{a}{m_{\varphi_1} m_{\varphi_2}} \left[m_{\zeta_1} \cos \theta_L \cos \theta_R F\left( \frac{m_{\varphi_1}^2}{m_{\zeta_1}^2}, \frac{m_{\varphi_2}^2}{m_{\zeta_1}^2} \right) \right.\\
        &\hspace{5cm}\left. + m_{\zeta_2} \sin \theta_L \sin \theta_R F\left( \frac{m_{\varphi_1}^2}{m_{\zeta_2}^2}, \frac{m_{\varphi_2}^2}{m_{\zeta_2}^2} \right) \right] \\
        m_b &= \frac{2 y_L^{Q_3} y_R^b}{16\pi^2} \frac{y_H v}{\sqrt{2}} \left[ \cos^2 \theta_s F\left( \frac{m_{\zeta_1}^2}{m_{\varphi_1}^2}, \frac{m_{\zeta_2}^2}{m_{\varphi_1}^2} \right) + \sin^2 \theta_s F\left( \frac{m_{\zeta_1}^2}{m_{\varphi_2}^2}, \frac{m_{\zeta_2}^2}{m_{\varphi_2}^2} \right) \right] \\
        m_\tau &= \frac{3 y_L^L y_R^\tau}{16\pi^2} \frac{y_H v}{\sqrt{2}} \left[ \cos^2 \theta_s F\left( \frac{m_{\zeta_1}^2}{m_{\varphi_1}^2}, \frac{m_{\zeta_2}^2}{m_{\varphi_1}^2} \right) + \sin^2 \theta_s F\left( \frac{m_{\zeta_1}^2}{m_{\varphi_2}^2}, \frac{m_{\zeta_2}^2}{m_{\varphi_2}^2} \right) \right]
    \end{split}
\end{equation}
where we have defined
\begin{equation}
    F(x,y) \equiv \frac{\sqrt{xy}}{x-y} \left(\frac{x}{x-1} \ln \left(x\right) - \frac{y}{y-1} \ln \left(y\right) \right)~.
\end{equation}
To calculate the effective Yukawa couplings, it is also helpful to define
\begin{equation}
    \begin{split}
        F(x) &\equiv \lim_{y\rightarrow x} F(x,y) = \frac{x}{x-1} - \frac{x}{(x-1)^2} \ln \left(x \right) \\ 
        G(x,y) &\equiv \frac{1}{x-y} \left( \frac{y(x+y)-2x}{(y-1)^2} y\ln \left(y \right) - \frac{x(x+y)-2y}{(x-1)^2} x \ln \left(x \right) \right)  + \frac{x}{x-1} +\frac{y}{y-1} \\
        Y_1(x,y) &\equiv \cos^2 2\theta_s F(x,y) + \frac{\sin^2 2\theta_s}{2}\left(\sqrt{\frac{y}{x}} F(x) + \sqrt{\frac{x}{y}} F(y) \right) \\
        Y_2(x,y) &\equiv F(x,y) + \frac{\sin 2\theta_L \sin 2\theta_R}{2} G(x,y)
    \end{split}
\end{equation}
The expressions for $Y_1$ and $Y_2$ correspond to the form that effective Yukawa couplings have in pure Class One and Class Two models respectively, in the limit that the Higgs boson four-momentum $p_h \rightarrow 0$.

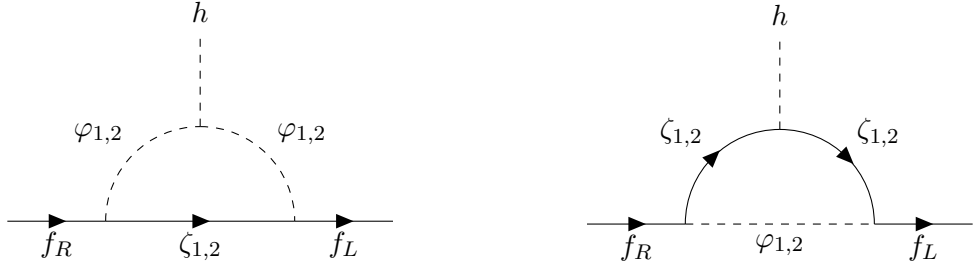
\begin{figure}[t]
    \centering
    \begin{subfigure}[t]{0.49\textwidth}
        \centering
        \begin{tikzpicture}
        \begin{feynman}
\vertex (a1);
\vertex[right=1.3cm of a1] (a2);
\vertex[right=2.5cm of a2] (a3);
\vertex[right=1.3cm of a3] (a4);
\vertex[right=1.25cm of a2] (fill);
\vertex[above=1.25 cm of fill] (d);

\vertex[above=1.25cm of d] (e) {\(h\)};

\diagram* {
(a1) -- [fermion, edge label'=$f_R$] (a2) -- [with arrow=0.5, edge label'=$\zeta_{1,2}$] (a3) -- [fermion, edge label'=$f_L$] (a4),
(a2) -- [scalar, quarter left, edge label=$\varphi_{1,2}$] (d) -- [scalar, quarter left, edge label=$\varphi_{1,2}$] (a3),
(d) -- [scalar] (e),
};

\end{feynman}
        \end{tikzpicture}
    \end{subfigure}%
    ~ 
    \begin{subfigure}[t]{0.49\textwidth}
        \centering
        \begin{tikzpicture}
        \begin{feynman}
\vertex (a1);
\vertex[right=1.3cm of a1] (a2);
\vertex[right=2.5cm of a2] (a3);
\vertex[right=1.3cm of a3] (a4);
\vertex[right=1.25cm of a2] (fill);
\vertex[above=1.25 cm of fill] (d);

\vertex[above=1.25cm of d] (e) {\(h\)};

\diagram* {
(a1) -- [fermion, edge label'=$f_R$] (a2) -- [scalar, edge label'=$\varphi_{1,2}$] (a3) -- [fermion, edge label'=$f_L$] (a4),
(a2) -- [with arrow=0.5, quarter left, edge label= $\zeta_{1,2}$] (d) -- [with arrow=0.5, quarter left, edge label = $\zeta_{1,2}$] (a3),
(d) -- [scalar] (e),
};

\end{feynman}
        \end{tikzpicture}
    \end{subfigure}
    \caption{Effective Yukawa coupling diagrams at one loop in the H2 model after EWSB. The couplings for all three SM fermions have contributions from both of these topologies.}
    \label{fig:h2_yukawas}
\end{figure}

As in all hybrid models, both the exotic fermions and scalars couple to the SM Higgs. The result is that the one-loop effective Yukawas for all relevant SM fermions will have contributions from both topologies in Figure~\ref{fig:h2_yukawas}, regardless of which topology a particular fermion's Yukawa diagram was before EWSB. The full expressions are remarkably unwieldy, so while we use them to produce our plots, here we only provide the expressions in certain limits. 
For $\theta_L \approx \theta_R$ and $p_h \rightarrow 0$, the charm quark effective Yukawa coupling is given by
\begin{equation}\label{eq:cyukawa}
    \begin{split}
        y_c^\mathrm{eff} = \frac{2 y_L^{Q_2} y_R^c}{16 \pi^2} &\left[ \frac{a}{m_{\varphi_1} m_{\varphi_2}} \left( m_{\zeta_1} \cos \theta_L \cos \theta_R Y_1\left( \frac{m_{\varphi_1}^2}{m_{\zeta_1}^2}, \frac{m_{\varphi_2}^2}{m_{\zeta_1}^2} \right) \right. \right. \\
        & \hspace{2cm} \left. + m_{\zeta_2} \sin \theta_L \sin \theta_R Y_1\left( \frac{m_{\varphi_1}^2}{m_{\zeta_2}^2}, \frac{m_{\varphi_2}^2}{m_{\zeta_2}^2} \right) \right) \\
        & \hspace{2cm} \left. + y_H \sin \theta_s \cos \theta_s \left(- Y_2\left(\frac{m_{\zeta_1}^2}{m_{\varphi_1}^2}, \frac{m_{\zeta_2}^2}{m_{\varphi_1}^2} \right) + Y_2\left(\frac{m_{\zeta_1}^2}{m_{\varphi_2}^2}, \frac{m_{\zeta_2}^2}{m_{\varphi_2}^2} \right) \right) \right]
    \end{split}
\end{equation}
The $b$ and $\tau$ effective Yukawas will differ slightly due to their original topologies being of a different class. For $m_\phi \approx m_\eta$ and $p_h\rightarrow 0$, the $b$ Yukawa coupling is
\begin{equation}\label{eq:byukawa}
\begin{split}
    y_b^\mathrm{eff} = &\frac{2y_L^{Q_3} y_R^b y_H}{16\pi^2} \left[\left( \cos^2 \theta_s Y_2\left(\frac{m_{\zeta_1}^2}{m_{\varphi_1}^2}, \frac{m_{\zeta_2}^2}{m_{\varphi_1}^2} \right) +\sin^2 \theta_s Y_2\left(\frac{m_{\zeta_1}^2}{m_{\varphi_2}^2}, \frac{m_{\zeta_2}^2}{m_{\varphi_2}^2} \right) \right) \right. \\
    & \qquad \left. + \frac{a}{m_{\varphi_1} m_{\varphi_2}} \frac{v}{\sqrt{2}} \frac{m_{\zeta_1} m_{\zeta_2}}{m_{\zeta_2}^2 - m_{\zeta_1}^2} \left(- Y_1\left(\frac{m_{\varphi_1}^2}{m_{\zeta_1}^2}, \frac{m_{\varphi_2}^2}{m_{\zeta_1}^2} \right) + Y_1\left(\frac{m_{\varphi_1}^2}{m_{\zeta_2}^2}, \frac{m_{\varphi_2}^2}{m_{\zeta_2}^2} \right)  \right) \right]
\end{split}
\end{equation}
The $\tau$ quark effective Yukawa will have the same form, and can be written as
\begin{equation}\label{eq:tauyukawa}
    y_\tau^\mathrm{eff} = \frac{3 y_L^L y_R^\tau}{2 y_L^{Q_3} y_R^b} y_b^\mathrm{eff} = \frac{m_\tau}{m_b} y_b^\mathrm{eff}
\end{equation}

These effective couplings possess a critical difference to those seen in pure models. In the latter, one finds that $y_f^\mathrm{eff} > y_f^\mathrm{SM}$ for all values of the model parameters, with equality approximately recovered as the scale of new physics is increased. However, analysis of the above expressions reveals that hybrid models should be able to produce $y_f^\mathrm{eff}$ greater than, equal to, \textit{and} less than $y_f^\mathrm{SM}$ in certain regions of parameter space, and indeed we find this to be the case in Section~\ref{sec:H2_pheno}. This difference stems from the fact that both types of topologies contribute to the effective Yukawa couplings in hybrid models, and these differing contributions can partially offset one another, whereas pure models will have contributions from only one type of topology. We may thus conclude that the predicted SM relationship $y_f = \sqrt{2} m_f / v$ can be replicated in radiative mass models, albeit this may come at the cost of significant tuning of the model parameters.

\subsection{Phenomenology}\label{sec:H2_pheno}

We examined several theoretical and experimental constraints on the H2 model, to see what the typical limitations on hybrid models in the catalogue may be. For simplicity we set the exotic fermion and scalar mass scales to be equal, $m_{\varphi_1} = m_{\zeta_1} \equiv m_1$. 

The first theoretical constraint we consider is unitarity. The expressions for $m_b$ and $m_\tau$ in eq.~\ref{eq:SMf_masses} can be rearranged to give expressions for the products $y_H y_L^{Q,L} y_R^{b,\tau}$. Since $0< F(x,y) \le 1$ and $y_{H, L, R} \le \sqrt{4\pi}$, these expressions give lower bounds $y_H \ge 0.151$ and $y_H \ge 0.045$ respectively. At values of $y_H$ close to these lower bounds, the unitarity constraints on the $y_L^{L,Q} y_R^{\tau,b}$ products will be especially stringent, since they are restricted to narrow ranges just below $4\pi$. Writing $y_L^{Q_2} y_R^c$ in terms of $m_c$ gives the strongest unitarity constraint on the remaining parameter space for larger $y_H$, so it is the only of these we will show explicitly on our plots.

We also consider the degree of fine-tuning needed to obtain the observed Higgs mass parameter $\mu_H^2$, adopting the fine-tuning measure defined in \cite{finetune}. In our plots we take $\Delta > 100$ to be a benchmark value for a high degree of fine-tuning, where regions of parameter space may be technically permitted but theoretically questionable. As is the theme for this section, both the exotic scalars and fermions will have non-zero contributions to $\mu_H^2$ at one loop.

The experimental constraints we consider are signal strengths for Higgs decays, corrections to electroweak physics, and the bound on the rare decay $t\rightarrow ch$. In the $\kappa$ framework~\cite{kappaframe,kappaframelong}, the signal strength for a process $ii \rightarrow h \rightarrow jj$ is defined as
\begin{equation}
    \mu_j^i = \frac{\kappa_i^2 \kappa_j^2}{\kappa_h^2}
\end{equation}
and normalised to unity at the SM prediction. Here $\kappa_{i,j}$ are measures of the strength of the $ii$ and $jj$ couplings to the Higgs relative to their SM values, so $\kappa_f = |y_f^\mathrm{eff} / y_f^\mathrm{SM}|$ for any fermion $f$, and $\kappa_h$ is a measure of the total Higgs width. The current constraints on the relevant $\mu_j$ are given in Table~\ref{tab:mu_current}. For the purposes of our analysis, we take
\begin{equation}
\begin{split}
    \mu_\tau = \frac{\kappa_g^2 \kappa_\tau^2}{\kappa_h^2} \quad &\text{and}\quad
    \mu_{b,c} = \frac{\kappa_{b,c}^2}{\kappa_h^2}\ , \\
    \mu_Z = \frac{\kappa_g^2}{\kappa_h^2} \quad &\text{and} \quad \mu_\gamma = \frac{\kappa_g^2 \kappa_\gamma^2}{\kappa_h^2}\ .
\end{split}
\end{equation}
This is justified since the dominant production mechanism contributing to the measurements of $\mu_{\tau,Z,\gamma}$ is gluon fusion, while the dominant contributions to $\mu_b$ and $\mu_c$ are from $VH$ production and $t\overline{t}$ annihilation, and corrections to the $hWW$, $hZZ$, and $h t\overline{t}$ couplings will be negligible. $\kappa_{b, c, \tau}$ and $\kappa_h$ are calculated using eqs.~\ref{eq:cyukawa} to \ref{eq:tauyukawa}, and we derive $\kappa_{g,\gamma}$ from the relevant expressions in an Appendix of~\cite{Baker_2021}. From these we can also see that the only way to obtain $\mu_{b,c} < 1$ is to have $\kappa_{b,c}^2 < \kappa_h^2$. Since $h\rightarrow b \overline{b}$ dominates the Higgs decay width, the only way to obtain $\mu_b < 1$ is thus to have $\kappa_b<1$. This is impossible in ``pure" models, but achievable in hybrid models which are capable of producing $y^\mathrm{eff} < y^\mathrm{SM}$.

\begin{table}[t]
\centering
\begin{tabular}{c|ccccc}
Process         & $h\rightarrow \tau^+ \tau^-$ & $h \rightarrow b \overline{b}$ & $h \rightarrow c \overline{c}$ & $h \rightarrow ZZ$ & $h \rightarrow \gamma \gamma$ \\ \hline
$\mu_j$ & $0.91 \pm 0.09$              & $0.94 \pm 0.11$                & $-0.5 \pm 3.4$  & $1.02 \pm 0.08$ & $1.10 \pm 0.06$ 
\end{tabular}
\caption{Current signal strengths for Higgs boson decays into $b,c,\tau, Z, \gamma$~\cite{ParticleDataGroup:2026aaa}. The primary contributions to these fits are from ATLAS and CMS data \cite{ATLAS_tau,ATLAS_bb,ATLAS_Z,ATLAS_gamma,CMS_sum,CMS_cc}.}
\label{tab:mu_current}
\end{table}

The introduction of the EW-charged exotics induces corrections to the vacuum polarisations of the EW gauge bosons, which are constrained by the Peskin-Takeuchi parameters $S, T$, and $U$~\cite{PTparamsearly,PTparams}. Contributions to $U$ are of formally higher order, and thus we only consider the modifications to $S$ and $T$. New physics contributions to these are highly constrained. Under the assumption that $U=0$, the current $1\sigma$ bounds on new contributions to $S$ and $T$ are~\cite{ParticleDataGroup:2026aaa}
\begin{equation}
        S = 0.008 \pm 0.071\quad \text{and}\quad 
        T = 0.021 \pm 0.055.
\end{equation}
In the H2 model, the respective one-loop scalar and fermion contributions to $T$ are
\begin{equation}
    \begin{split}
        T_S = \frac{3}{16 \pi s_W^2 c_W^2 m_Z^2} &\left[ m_\eta^2 + c_s^2 M_{\varphi_1}^2 + s_s^2 m_{\varphi_2}^2 + \frac{2 c_s^2 m_\eta^2 m_{\varphi_1}^2}{m_\eta^2 - m_{\varphi_1}^2} \ln \left(\frac{m_{\varphi_1}^2}{m_\eta^2} \right) \right. \\
        & \qquad \left. +\frac{2 s_s^2 m_\eta^2 m_{\varphi_2}^2}{m_\eta^2 - m_{\varphi_2}^2} \ln \left(\frac{m_{\varphi_2}^2}{m_\eta^2} \right) + \frac{2 c_s^2 s_s^2 m_{\varphi_1}^2 m_{\varphi_2}^2}{m_{\varphi_2}^2 - m_{\varphi_1}^2} \ln \left(\frac{m_{\varphi_2}^2}{m_{\varphi_1}^2} \right) \right] 
    \end{split}
\end{equation}
\begin{equation}
    \begin{split}
        T_{F} = \frac{3}{16 \pi s_W^2 c_W^2 m_Z^2} \Biggl[8 s_L s_R c_L c_R m_{\zeta_1} m_{\zeta_2} + m_{\zeta_1}^2(c_L^4 + c_R^4) + m_{\zeta_2}^2 (s_L^4 + s_R^4) - 6 m_\psi^2  \\
        + \left( \frac{3 s_L^2 c_R^2 - s_R^2 c_L^2}{m_{\zeta_2}^2 - m_{\zeta_1}^2}  - \frac{c_L^2 - 3c_R^2}{m_{\zeta_1}^2 - m_\psi^2}\right) 2 m_{\zeta_1}^4 \ln \left( \frac{m_{\zeta_1}^2}{m_\psi^2} \right) \\
        + \left( \frac{3 c_L^2 s_R^2 - c_R^2 s_L^2}{m_{\zeta_1}^2 - m_{\zeta_2}^2}  - \frac{s_L^2 - 3s_R^2}{m_{\zeta_2}^2 - m_\psi^2}\right) 2 m_{\zeta_2}^4 \ln \left( \frac{m_{\zeta_2}^2}{m_\psi^2} \right) \\
        +\left.  \frac{8 m_{\zeta_1}^2 m_{\zeta_2}^2 (s_L^2 - s_R^2)^2}{m_{\zeta_1}^2 - m_{\zeta_2}^2} \ln \left( \frac{m_{\zeta_2}^2}{m_{\zeta_1}^2} \right) \right]
    \end{split}
\end{equation}
where we have denoted $s_i = \sin \theta_i$ and $c_i = \cos \theta_i$, and $\theta_W$ is the Weinberg angle. The new contributions to $S$ are too lengthy to reproduce here in full, so we instead provide them in certain limits. For small $\theta_s$ (corresponding to $m_\eta^2 > m_\phi^2 \gg \frac{av}{\sqrt{2}}$) and $m_{\varphi_1} \gg m_Z$, the scalar contribution to $S$ is
\begin{equation}
\begin{split}
    S_S = \frac{a^2 v^2}{24 \pi m_\eta^2 (m_\eta^2 - m_\phi^2)^5} \Bigg[&m_\phi^2 (5 m_\phi^6 - 15 m_\phi^4 m_\eta^2 + 3 m_\phi^2 m_\eta^4 +7 m_\eta^6) \\
    &\left.+ 2(4m_\phi^6 m_\eta^2 - 12 m_\phi^4 m_\eta^4 + 3 m_\phi^2 m_\eta^6 - m_\eta^8) \ln \left( \frac{m_\eta^2}{m_\phi^2} \right) \right]
\end{split}
\end{equation}
For small $\theta_{L,R}$ (corresponding to $m_\chi > m_\psi \gg \frac{y_H v}{\sqrt{2}}$) and $m_{\zeta_1} \gg m_Z$, the fermion contribution to $S$ is
\begin{equation}
\begin{split}
    S_F = \frac{y_H^2 v^2}{6\pi (m_\chi^2 - m_\psi^2)^5} \Bigg[&m_\psi^2(2m_\psi^6 - 3 m_\psi^4 m_\chi^2 + 6 m_\psi^2 m_\chi^4 -5 m_\chi^6) \\
    &\left. + (2 m_\psi^8 - 4m_\psi^6 m_\chi^4 + 9 m_\psi^4 m_\chi^4 - 2 m_\psi^4 m_\chi^6 + m_\chi^8) \ln \left(\frac{m_\chi^2}{m_\psi^2} \right)  \right]
\end{split}
\end{equation}

The exotic couplings to $Q_L^3$ and $c_R$ will induce mixing between the top and charm quarks. The effective Lagrangian for the relevant lowest order interactions is
\begin{equation}
    -\lagr_\mathrm{eff} \supset \mqty(\overline{t_L} & \overline{c_L}) \left[ \mqty(\frac{y_t v}{\sqrt{2}} & \frac{y_L^{Q_3}}{y_L^{Q_2}} m_c \\ \frac{(y_L^{Q_3} y_R^c)^*}{y_L^{Q_2} y_R^c} m_c & m_c) + \mqty(y_t& \frac{y_L^{Q_3}}{y_L^{Q_2}} y_c^\mathrm{eff} \\ \frac{(y_L^{Q_3} y_R^c)^*}{y_L^{Q_2} y_R^c} y_c^\mathrm{eff} & y_c^\mathrm{eff}) \frac{h}{\sqrt{2}} \right] \mqty(t_R \\ c_R) + \hc 
\end{equation}
where $m_c$ and $y_c^\mathrm{eff}$ are as defined in eqs~\ref{eq:SMf_masses} and~\ref{eq:cyukawa}.
In general these two matrices will not be simultaneously diagonalisable. Thus when the effective Lagrangian is rewritten in terms of the mass basis, the effective Yukawa matrix will have off diagonal elements $y_{tc}$ which induce the rare decay $t \rightarrow ch$. The values of these off-diagonal elements depend on $y_c^\mathrm{eff}$, $y_L^{Q_{2,3}}$, and $y_R^c$. Since there is some freedom in how the $y_L$ and $y_R$ vary with respect to one another, we take $y_L^{Q_{2,3}} = y_R^{b,c}$ for simplicity. We can then rearrange eq.~\ref{eq:SMf_masses} to find expressions for $y_L^{Q_{2,3}}$ and $y_R^c$ in terms of $m_c$, $m_b$, and the other model parameters, so as to write the $y_{tc}$ as a function of these model parameters. We refrain from displaying the result here, since diagonalising into the mass basis is a trivial computation that nonetheless produces an extremely complicated expression for these off-diagonal entries. The $t\rightarrow ch$ partial decay width can be written
\begin{equation}
    \Gamma_{t\rightarrow ch} = \frac{m_t}{32\pi} \left(1 - \frac{m_h^2}{m_t^2} \right) y_{tc}^2
\end{equation}
since $m_c \ll m_h, m_t$. If we take $Br(t\rightarrow bW)$ to be approximately $1$, we can estimate the branching fraction as
\begin{equation}
    Br(t\rightarrow ch) \simeq \frac{\Gamma_{t\rightarrow ch}}{\Gamma_{t\rightarrow bW}} = 0.261 \cdot y_{tc}^2
\end{equation}
using observed values for the various masses. Currently the $2\sigma$ upper bound on $t\rightarrow ch$ is $3.4 \cdot 10^{-4}$, though next generation experiments could bring it as low as $1.6\cdot 10^{-5}$~\cite{ESUbriefing}. 

\begin{figure}[t]
    \centering
    \begin{subfigure}[b]{0.43\textwidth}
        \centering
        \includegraphics[width=0.99\textwidth]{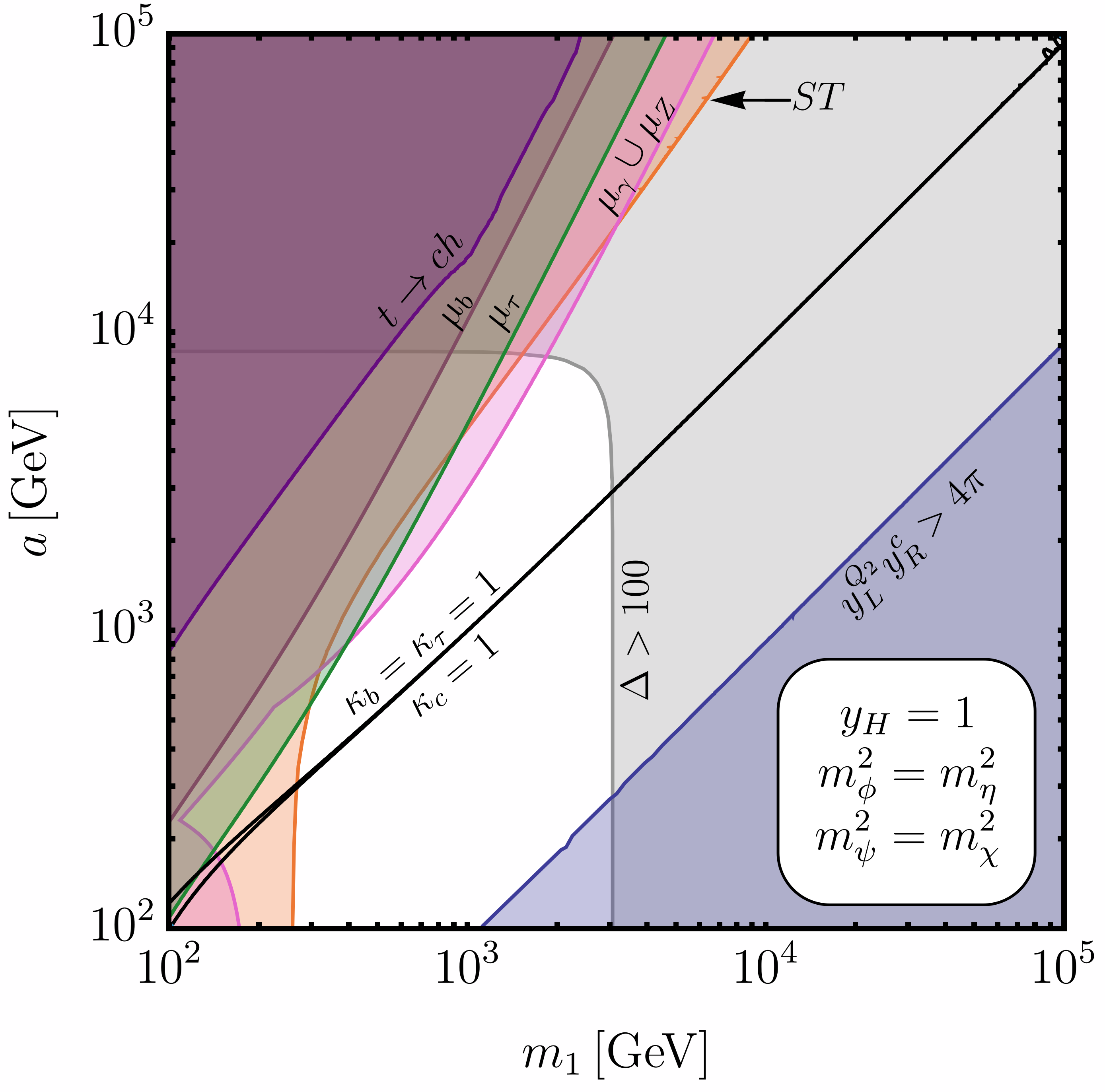}
    \end{subfigure}%
    ~ 
    \begin{subfigure}[b]{0.43\textwidth}
        \centering
        \includegraphics[width=0.99\textwidth]{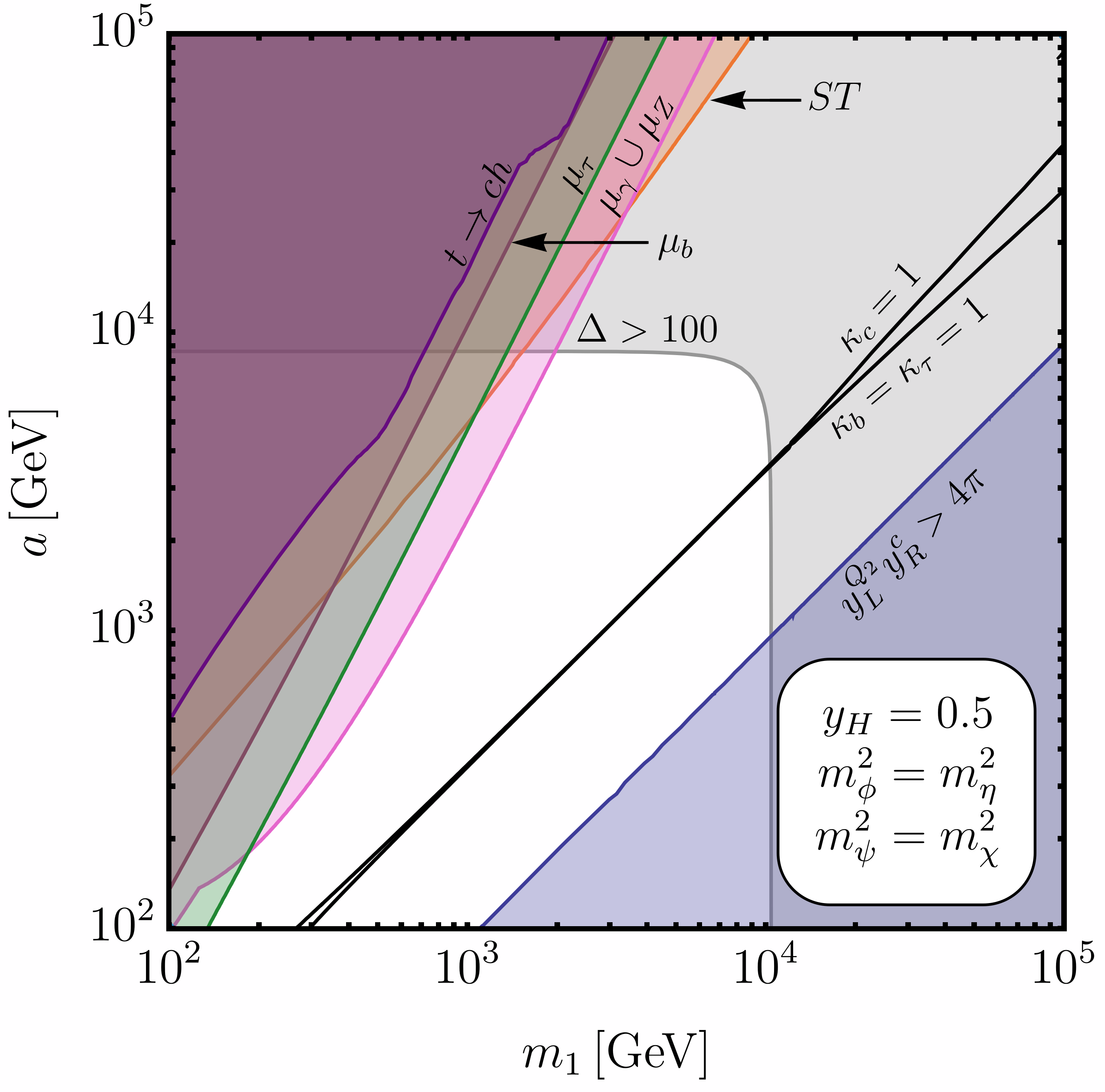}
    \end{subfigure}%
    
    \begin{subfigure}[b]{0.43\textwidth}
        \centering
        \includegraphics[width=0.99\textwidth]{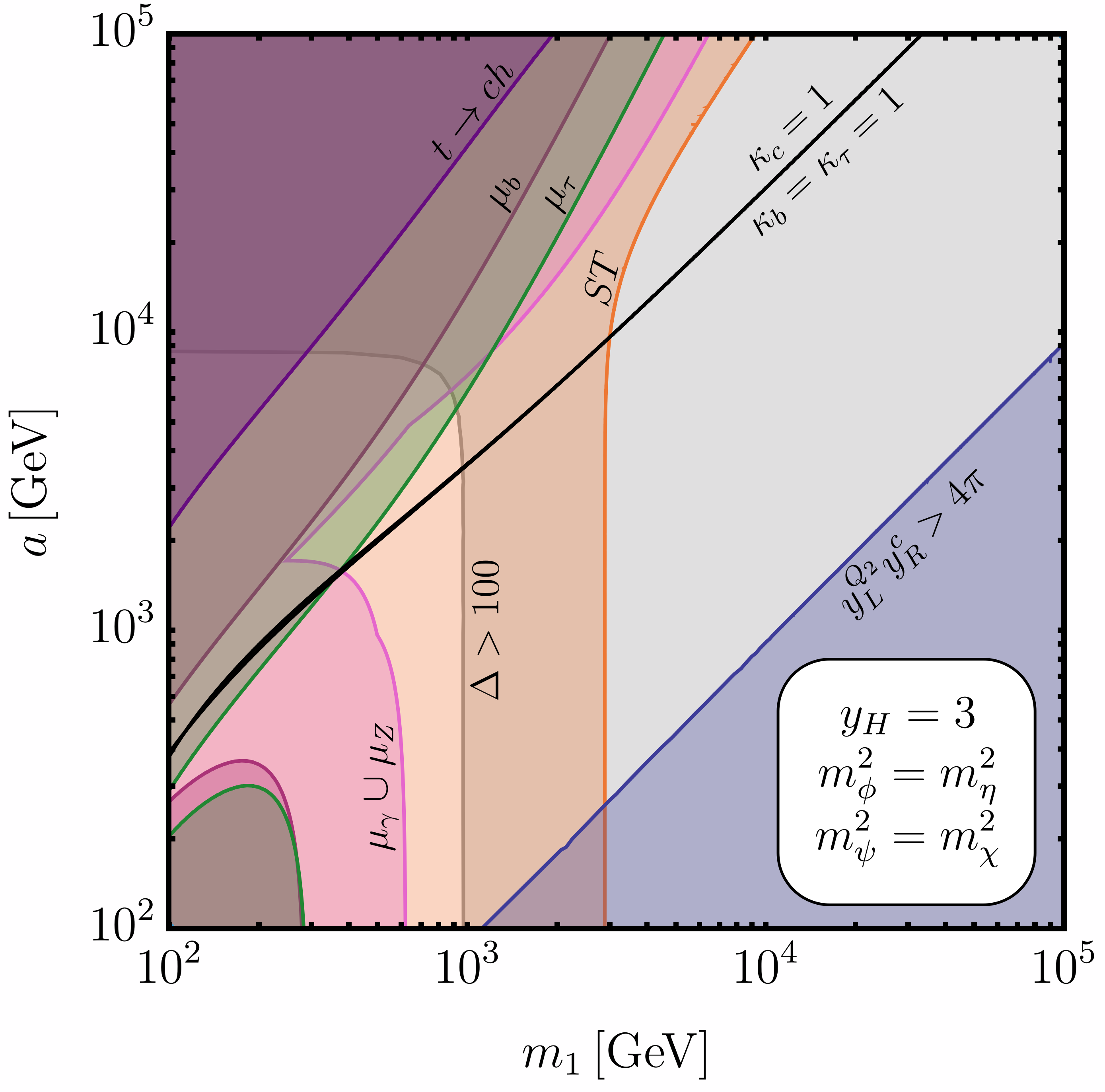}
    \end{subfigure}%
    ~ 
    \begin{subfigure}[b]{0.43\textwidth}
        \centering
        \includegraphics[width=0.99\textwidth]{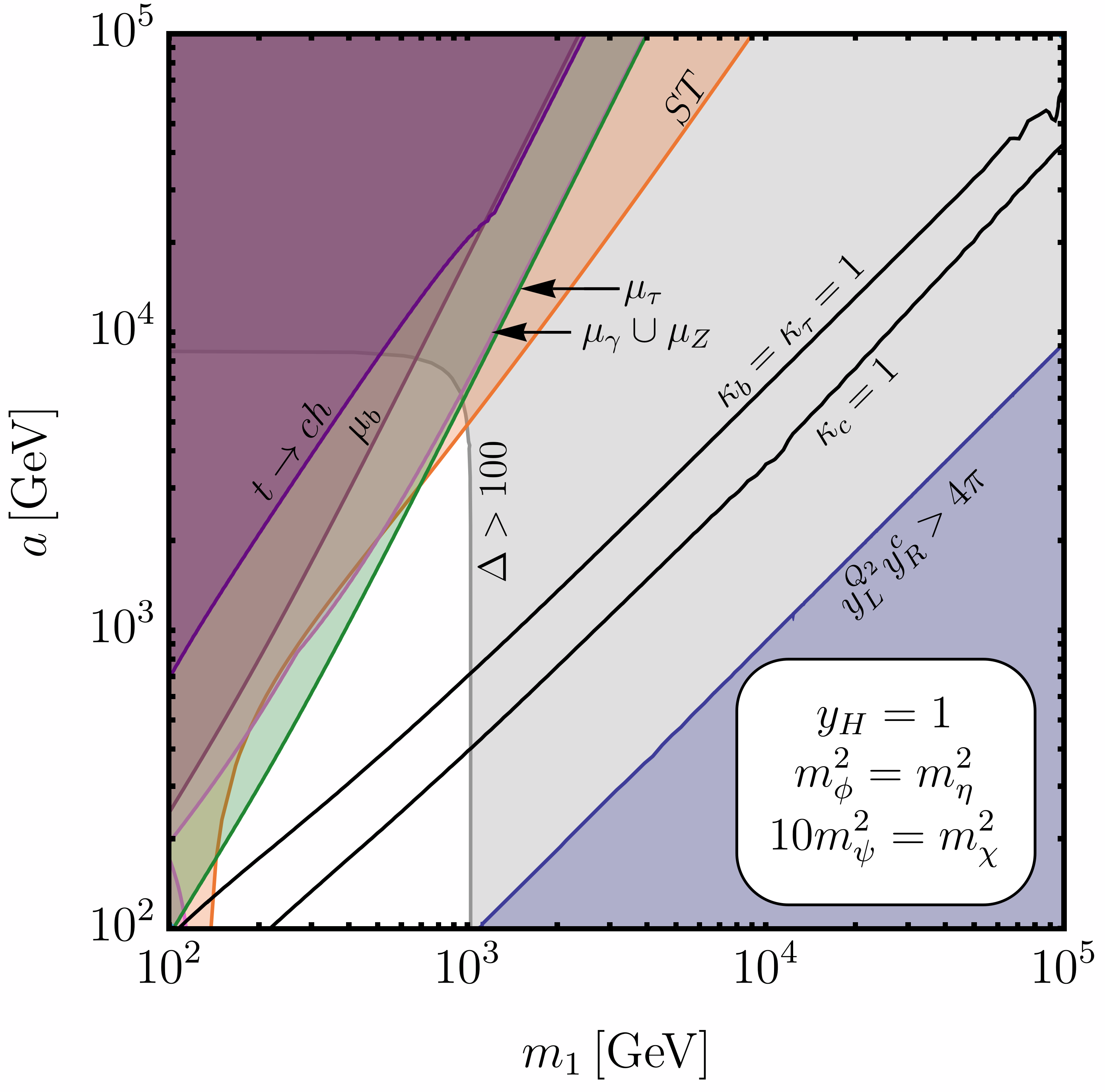}
    \end{subfigure}%
    \caption{Theoretical and 2$\sigma$ experimental constraints on the H2 model. Shaded regions are excluded, excepting $\Delta > 100$ where they are disfavoured. Parameter values which produce both $\kappa_b = \kappa_\tau=1$ and $\kappa_c = 1$ cannot be ruled out by improving constraints on Yukawa couplings. }
    \label{fig:h2_plots}
\end{figure}

\begin{figure}[t]
    \centering
    \begin{subfigure}[b]{0.43\textwidth}
        \centering
        \includegraphics[width=0.99\textwidth]{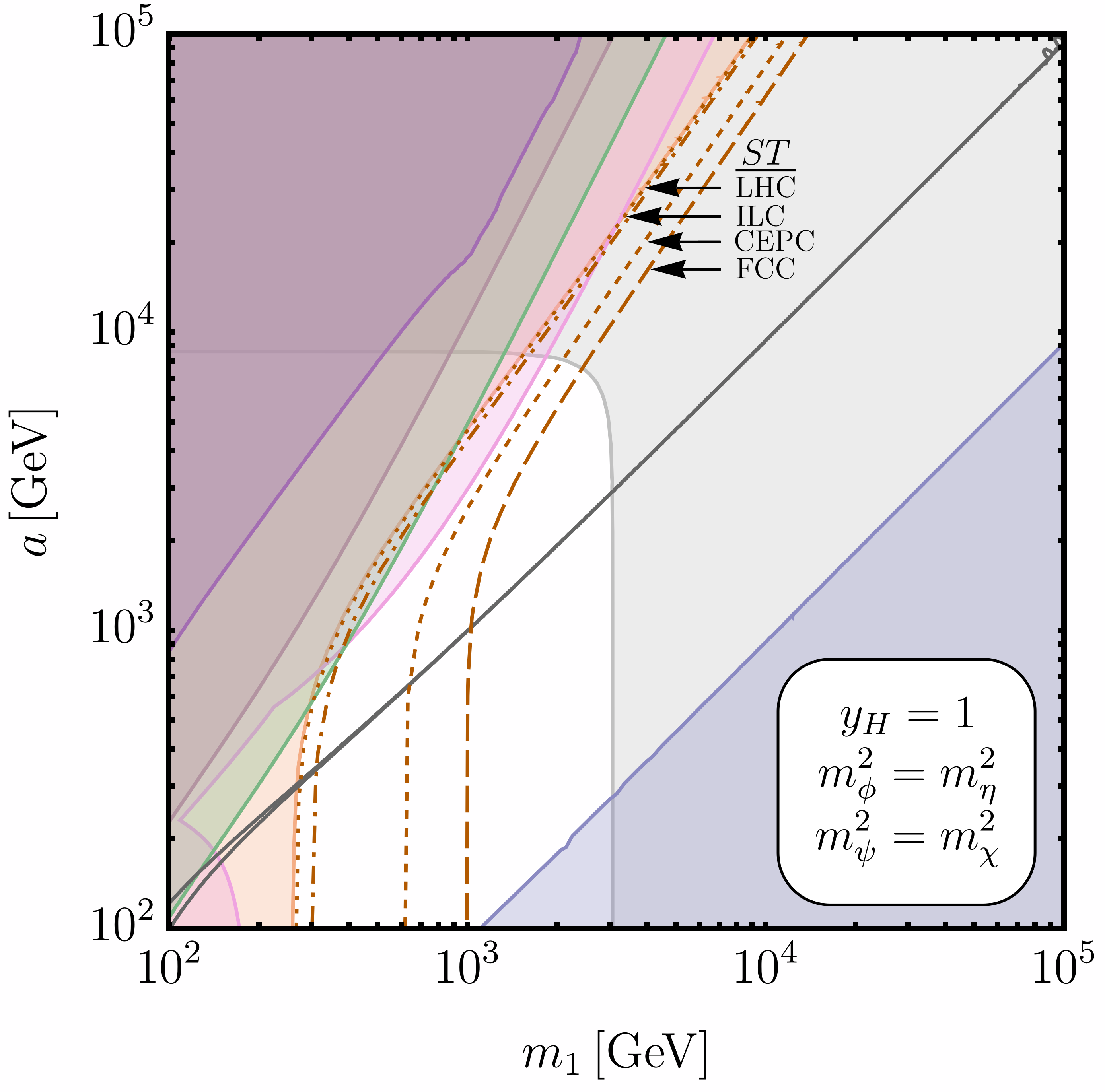}
    \end{subfigure}%
    ~ 
    \begin{subfigure}[b]{0.43\textwidth}
        \centering
        \includegraphics[width=0.99\textwidth]{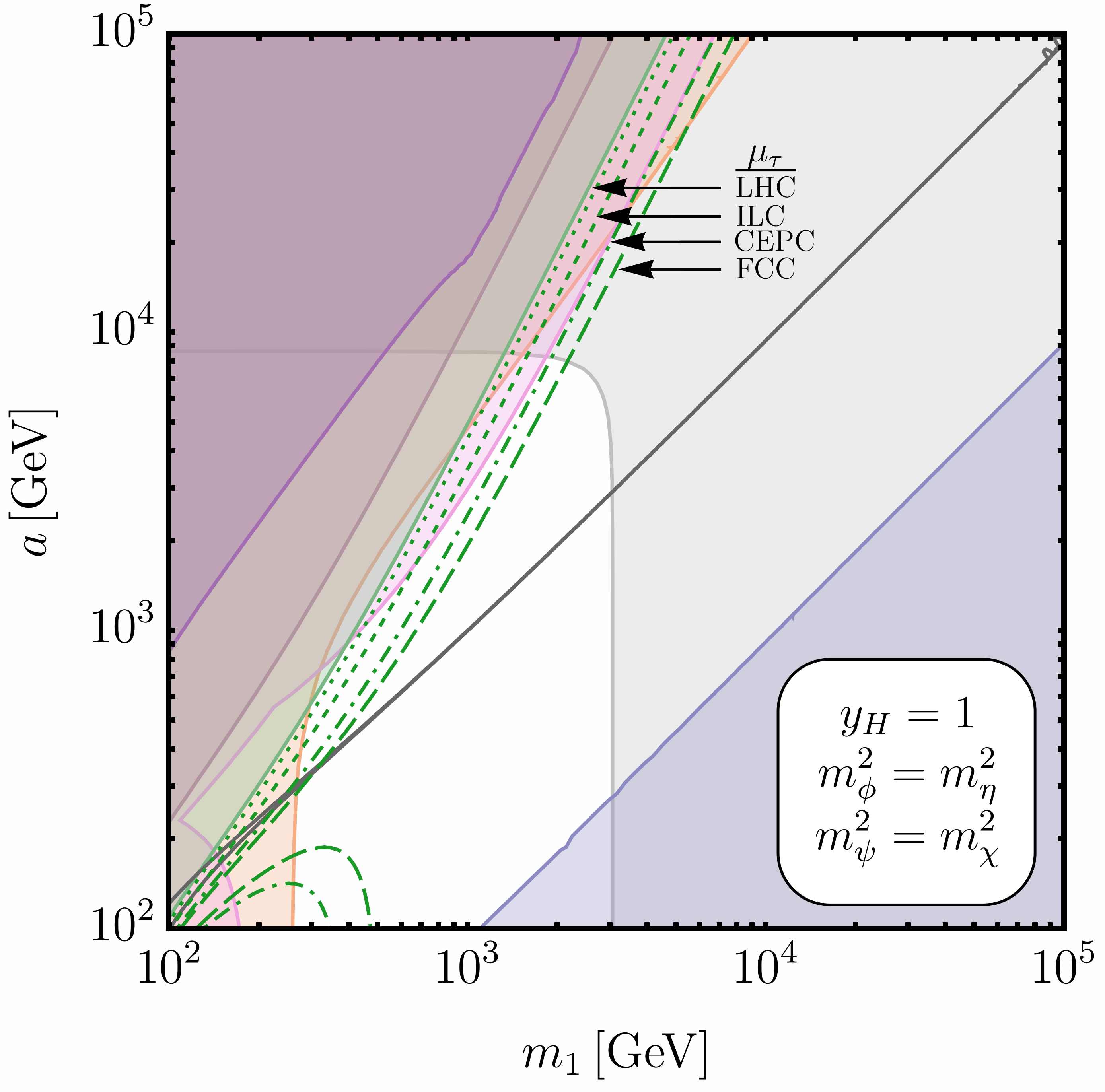}
    \end{subfigure}%
    
    \begin{subfigure}[b]{0.43\textwidth}
        \centering
        \includegraphics[width=0.99\textwidth]{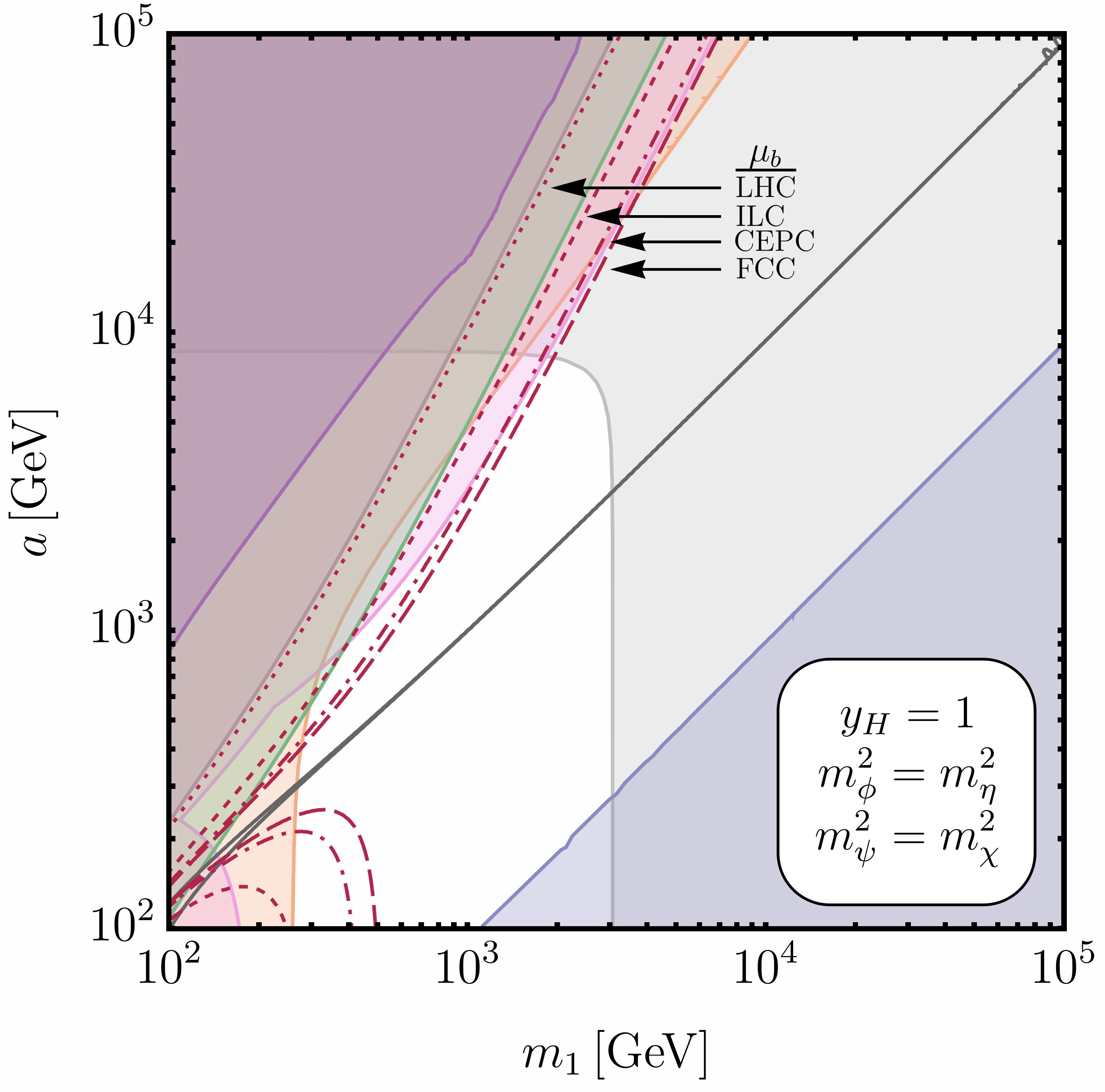}
    \end{subfigure}%
    ~ 
    \begin{subfigure}[b]{0.43\textwidth}
        \centering
        \includegraphics[width=0.99\textwidth]{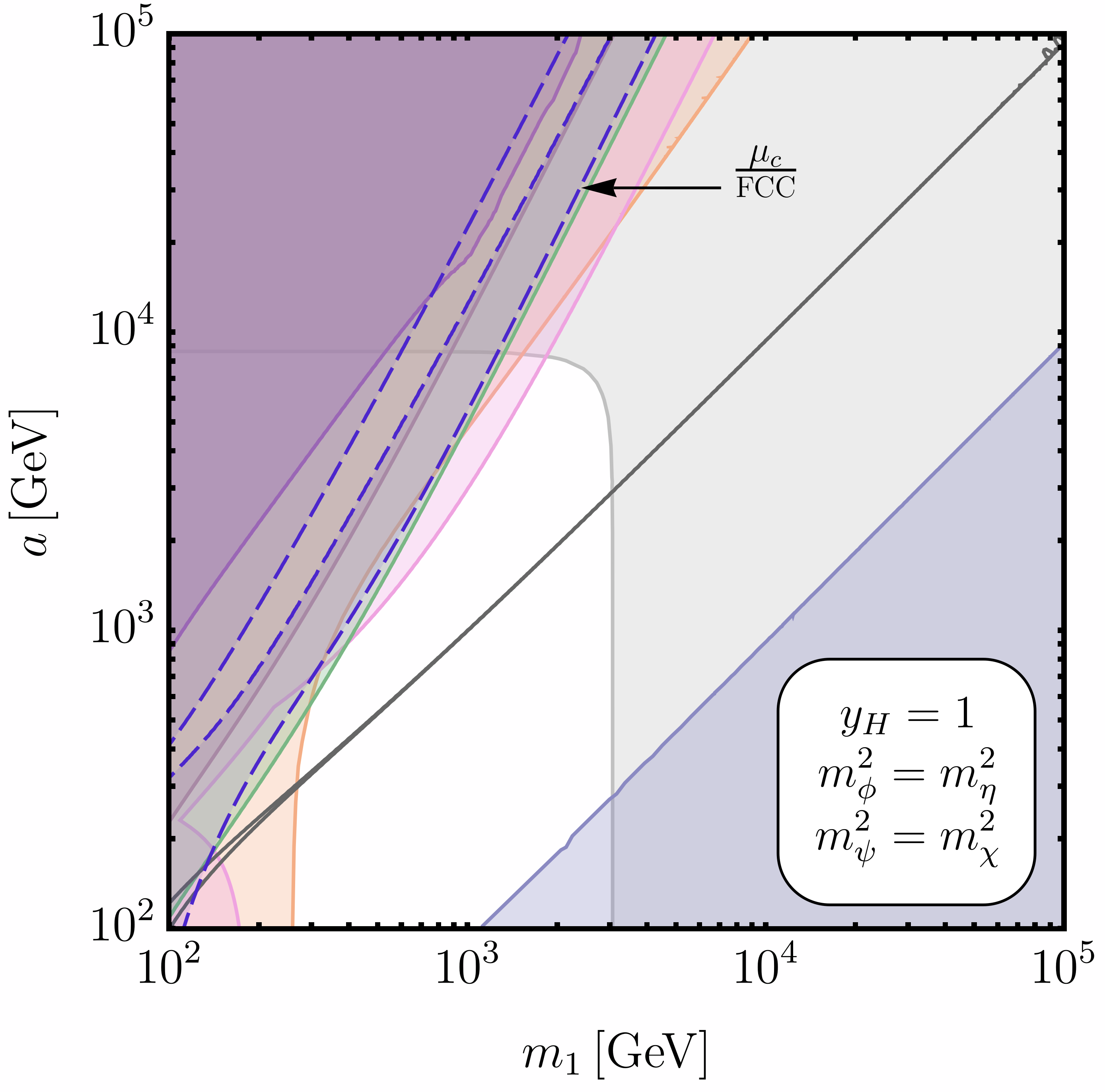}
    \end{subfigure}%
    \caption{Projected 2$\sigma$ experimental constraints on the H2 model, assuming central values at the SM prediction. Shaded regions correspond to the same constraints in Figure~\ref{fig:h2_plots}. Values for $\Delta \kappa_{b,c,\tau}$ are taken from~\cite{collider_future}, and $\Delta S$ and $\Delta T$ from~\cite{STfuture}. For clarity, only the FCC constraint on $\mu_c$ is shown; CEPC and ILC also produce projected bounds although they are weaker.}
    \label{fig:h2_futures}
\end{figure}

Exclusion plots using these constraints are shown for various slices of parameter space in Figure~\ref{fig:h2_plots}. The $\mu_\gamma$ and $\mu_Z$ bounds are combined for visual clarity, though $\mu_\gamma$ is typically stronger. Constraints from projected future sensitivities are shown in Figure~\ref{fig:h2_futures}, assuming central values for each measurement at the SM prediction. Projected $t\rightarrow ch$ bounds are weaker than the current experimental constraints, and so are not shown. 

The existence of unshaded regions demonstrates that viable parameter space currently exists for this model without excessive fine-tuning. Additionally, there are points within these regions with $\kappa_b = \kappa_\tau = \kappa_c = 1$ that are thus impossible to rule out by improving precision on $\kappa_{b,c,\tau}$ while agreement with the SM is maintained. To exclude these points, one would need more precise bounds on $S$ and $T$, or to carefully recast bounds from direct searches at colliders. Alternatively, such a model may be able to account for a small disparity between signal strength bounds and SM prediction, were one to arise in the future. Previous models in the framework could account for signal strengths slightly above unity, whereas hybrid models could account for measurements slightly below unity as well.

Of the experimental constraints, the $ST$ parameters and $\mu_{\gamma}$ typically provide the most stringent bounds for this model, including in future projections. There are multiple reasons for this. First, minimal hybrid models see larger corrections to these than previously considered models, as there is both a colour factor enhancement and contributions from both exotic fermion and scalar sectors. Second, the hybrid form of the Yukawa couplings can bring them into closer alignment with SM predictions, thus weakening the corresponding fermion signal strength bounds. Finally, the current 2$\sigma$ lower bound on $\mu_\gamma$ is 0.98, thus small decreases in $\kappa_\gamma$ or increases in $\kappa_h$ can quickly bring the predicted $\mu_\gamma$ into conflict with experiment.

Since the exotics are charged under $SU(3)_C$, we expect direct searches at colliders to provide significant constraints. As we have taken $\mathcal{Z}_2$ to be either unbroken or only broken by terms with small couplings, the lightest exotic field should be collider stable. The relevant searches are for $R$-hadrons, as long-lived coloured exotics will hadronise with coloured SM fields to form at least quasi-stable particles. If $m_{\varphi_1}$ is the lightest exotic, the existing bound on long lived sbottoms would enforce $m_{\varphi_1} > 1.25 \cdot 10^3$ GeV~\cite{SUSYbound}. We would expect a similar bound on $m_{\zeta_1}$ from recast searches if it was instead the lightest exotic. These bounds exclude a significant portion of the available parameter space not ruled out by other constraints or subject to a high degree of fine-tuning. Future experiments could see up to a $\sim$10x improvement to this bound~\cite{deBlas:2025gyz}, which could drastically reduce the viable parameter space without a fine-tuning problem, or remove it altogether. Since all hybrid models in the catalogue contain only coloured exotics, we would expect similar limitations on their parameter spaces.

\section{Conclusion}\label{sec:conclusion}
We have detailed an extension to an earlier framework for generating one-loop SM fermion masses and Yukawa couplings. We then classified all minimal models capable of generating one-loop masses for the bottom, charm, and tau within this framework, to explain why their masses are near the GeV level. We found there exist 25 such models, eight of which are ``hybrid" models; these have multiple topologies for the effective Yukawa coupling diagrams, and had not been previously studied in this framework. We analysed one of these hybrid models as a benchmark, and found that there currently exists parameter space where it is not ruled out or subject to a high degree of fine-tuning. We also showed that due to its hybrid nature, it was capable of producing effective Yukawa couplings greater than, less than, and exactly equal to the corresponding SM Yukawa couplings, demonstrating that the SM tree level mass-Yukawa coupling relationship can be produced by radiative mass models. If future measurements are found to be in tension with the SM relationship, such models may also be capable of explaining a discrepancy in either direction. It must be emphasised that even if agreement is maintained, these models will not necessarily be ruled out, but will be restricted to regions of parameter space with greater amounts of fine-tuning.
Other individual models in the catalogue also have unique features that may be worth exploring, such as the viability of dark matter candidates, mass-mixing between three or more exotic fields, and/or the potential for an axion to be introduced.  Most importantly, future work is needed to construct a complete model generating all SM fermion masses at the appropriate orders, using the catalogue here as a basis for further model-building. Flavour phenomenology for the lighter SM fermions is more tightly constrained, and may provide the strongest bounds on a complete model, though we expect that these would predominantly restrict the new parameters introduced to construct such a theory.

\acknowledgments
We thank Peter Cox for helpful discussions. This work was supported in part by the Australian Research Council through the ARC Discovery Project grant DP260104148. LS is supported by an Australian Government Research Training Program Scholarship and the Rowden White Scholarship.

\appendix

\section{Three-exotic models}\label{sec:three_exotics}
Let $\mathcal{A},\mathcal{B},\mathcal{C}$ be distinct chiral fermions with right-handed components that do not transform under $SU(2)_L$. Consider trying to generate all three of their masses at one-loop level within this framework, while introducing only three exotic fields. Without loss of generality, we can let these three fields be $\exlab_1^\mathcal{A}, \exlab_2^\mathcal{A}, \exlab_3^\mathcal{A}$.

Now we note that $\exlab_1^\mathcal{A}$ and $\exlab_2^\mathcal{A}$ have the same spin, which is different from that of $\exlab_3^\mathcal{A}$. Since this is also a property we require of $\exlab_{1,2,3}^\mathcal{B}$ and $\exlab_{1,2,3}^\mathcal{C}$, it must be that $\exlab_3^\mathcal{B}$ and $\exlab_3^\mathcal{C}$ are either the same as or the conjugate of $\exlab_3^\mathcal{A}$. Thus
\begin{equation}
    Y(\exlab_3^\mathcal{B}), Y(\exlab_3^\mathcal{C}) \in \{+Y(\exlab_3^\mathcal{A}), -Y(\exlab_3^\mathcal{A}) \}\ .
\end{equation}

Similarly, $\exlab_1^\mathcal{A}$ and $\exlab_3^\mathcal{A}$ transform the same way under $SU(2)_L$, and differently to how $\exlab_2^\mathcal{A}$ transforms. By analogous argument it is clear that $\exlab_2^\mathcal{B}$ and $\exlab_2^\mathcal{C}$ are the same as or the conjugate of $\exlab_2^\mathcal{A}$. Finally, this means that $\exlab_1^\mathcal{B}$ and $\exlab_1^\mathcal{C}$ must be identified with $\exlab_1^\mathcal{A}$ or its conjugate. Therefore
\begin{equation}
    Y(\exlab_1^\mathcal{B}), Y(\exlab_1^\mathcal{C}) \in \{+Y(\exlab_1^\mathcal{A}), -Y(\exlab_1^\mathcal{A}) \}\ .
\end{equation}

The term that couples $\mathcal{A}_R$ to $\exlab_1^\mathcal{A}$ and $\exlab_3^\mathcal{A}$ requires that $Y(\mathcal{A}_R) = Y(\exlab_1^\mathcal{A}) + Y(\exlab_3^\mathcal{A})$, and similarly for $\mathcal{B}$ and $\mathcal{C}$. From the above logic, we have that
 \begin{equation}
     |Y(\mathcal{B}_R)| = |Y(\exlab_1^\mathcal{B}) + Y(\exlab_3^\mathcal{B})| \in \{|Y(\exlab_1^\mathcal{A}) + Y(\exlab_3^\mathcal{A})|, |Y(\exlab_1^\mathcal{A}) - Y(\exlab_3^\mathcal{A})|\}\ ,
 \end{equation}
likewise $|Y(\mathcal{C}_R)|$ must be in this set.

So
\begin{equation}
    \begin{split}
        &\{|Y(\mathcal{A}_R)|, |Y(\mathcal{B}_R)|, |Y(\mathcal{C}_R)|\} \subseteq \{|Y(\exlab_1^\mathcal{A}) + Y(\exlab_3^\mathcal{A})|, |Y(\exlab_1^\mathcal{A}) - Y(\exlab_3^\mathcal{A})|\} \\
        &\Rightarrow |\{|Y(\mathcal{A}_R)|, |Y(\mathcal{B}_R)|, |Y(\mathcal{C}_R)|\}| \leq 2\ .
    \end{split}
\end{equation}
We therefore conclude that since $|\{|Y(\tau_R)|, |Y(b_R)|, |Y(c_R)|\}| = |\{1, \frac{1}{3}, \frac{2}{3} \}| = 3$, it is impossible to produce a three-exotic model in our framework for $b+c+\tau$.

\bibliographystyle{jhepbst}
\bibliography{References}{}

@article{Baker_2021,
    author = "Baker, Michael J. and Cox, Peter and Volkas, Raymond R.",
    title = "{Has the Origin of the Third-Family Fermion Masses been Determined?}",
    eprint = "2012.10458",
    archivePrefix = "arXiv",
    primaryClass = "hep-ph",
    doi = "10.1007/JHEP04(2021)151",
    journal = "JHEP",
    volume = "04",
    pages = "151",
    year = "2021"
}

@article{Carrasco-Martinez:2026wzu,
    author = "Carrasco-Martinez, Juanca and Hall, Lawrence J.",
    title = "{Maximal Abelian Flavor Symmetries}",
    eprint = "2606.07799",
    archivePrefix = "arXiv",
    primaryClass = "hep-ph",
    month = "6",
    year = "2026"
}

@article{Davidson:1987mh,
    author = "Davidson, Aharon and Wali, Kameshwar C.",
    title = "{Universal Seesaw Mechanism?}",
    reportNumber = "SU-4228-363",
    doi = "10.1103/PhysRevLett.59.393",
    journal = "Phys. Rev. Lett.",
    volume = "59",
    pages = "393",
    year = "1987"
}

@article{Davidson:1987tr,
    author = "Davidson, Aharon and Wali, Kameshwar C.",
    title = "{Family Mass Hierarchy From Universal Seesaw Mechanism}",
    reportNumber = "SU-4228-375",
    doi = "10.1103/PhysRevLett.60.1813",
    journal = "Phys. Rev. Lett.",
    volume = "60",
    pages = "1813",
    year = "1988"
}

@article{Davidson:1989bx,
    author = "Davidson, Aharon and Ranfone, Stefano and Wali, Kameshwar C.",
    title = "{Quark Masses and Mixing Angles From Universal Seesaw Mechanism}",
    reportNumber = "SU-4228-410",
    doi = "10.1103/PhysRevD.41.208",
    journal = "Phys. Rev. D",
    volume = "41",
    pages = "208",
    year = "1990"
}

@article{Davidson:1993xn,
    author = "Davidson, Aharon and Michel, Louis and Sage, Martin L. and Wali, Kameshwar C.",
    title = "{Quark mass hierarchies from universal seesaw mechanism}",
    reportNumber = "SU-4240-510",
    doi = "10.1103/PhysRevD.49.1378",
    journal = "Phys. Rev. D",
    volume = "49",
    pages = "1378--1388",
    year = "1994"
}

@article{Davidson:1998vr,
    author = "Davidson, Aharon and Schwartz, Tomer and Volkas, Raymond R.",
    title = "{Linking geometric mass hierarchy with threefold family replication}",
    eprint = "hep-ph/9802235",
    archivePrefix = "arXiv",
    reportNumber = "UMP-97-35, RCHEP-97-06, BGU-97-12",
    doi = "10.1088/0954-3899/25/8/302",
    journal = "J. Phys. G",
    volume = "25",
    pages = "1571--1588",
    year = "1999"
}

@article{Fraser_2014,
    author = "Fraser, Sean and Ma, Ernest",
    title = "{Anomalous Higgs Yukawa Couplings}",
    eprint = "1402.6415",
    archivePrefix = "arXiv",
    primaryClass = "hep-ph",
    reportNumber = "UCRHEP-T541-(FEB-2014), UCRHEP-T541-(MAR-2014), UCRHEP-T541-(JUL-2014), UCRHEP-T541-(SEP-2014)",
    doi = "10.1209/0295-5075/108/11002",
    journal = "EPL",
    volume = "108",
    number = "1",
    pages = "11002",
    year = "2014"
}

@article{Froggatt:1978nt,
    author = "Froggatt, C. D. and Nielsen, Holger Bech",
    title = "{Hierarchy of Quark Masses, Cabibbo Angles and CP Violation}",
    reportNumber = "CERN-TH-2519",
    doi = "10.1016/0550-3213(79)90316-X",
    journal = "Nucl. Phys. B",
    volume = "147",
    pages = "277--298",
    year = "1979"
}

@article{Hall:1995es,
    author = "Hall, Lawrence J. and Murayama, Hitoshi",
    title = "{A Geometry of the generations}",
    eprint = "hep-ph/9508296",
    archivePrefix = "arXiv",
    reportNumber = "UCB-PTH-95-29, LBL-37627",
    doi = "10.1103/PhysRevLett.75.3985",
    journal = "Phys. Rev. Lett.",
    volume = "75",
    pages = "3985--3988",
    year = "1995"
}

@article{King:2001uz,
    author = "King, S. F. and Ross, Graham G.",
    title = "{Fermion masses and mixing angles from SU(3) family symmetry}",
    eprint = "hep-ph/0108112",
    archivePrefix = "arXiv",
    reportNumber = "SHEP-01-21, OUTP-01-46P",
    doi = "10.1016/S0370-2693(01)01139-X",
    journal = "Phys. Lett. B",
    volume = "520",
    pages = "243--253",
    year = "2001"
}

@article{Pomarol:1995xc,
    author = "Pomarol, Alex and Tommasini, Daniele",
    title = "{Horizontal symmetries for the supersymmetric flavor problem}",
    eprint = "hep-ph/9507462",
    archivePrefix = "arXiv",
    reportNumber = "CERN-TH-95-207",
    doi = "10.1016/0550-3213(96)00074-0",
    journal = "Nucl. Phys. B",
    volume = "466",
    pages = "3--24",
    year = "1996"
}

@article{Rajpoot:1986nv,
    author = "Rajpoot, S.",
    title = "{Seesaw Masses for Quarks and Leptons in an Ambidextrous Electroweak Interaction Model}",
    reportNumber = "Print-86-1477 (OKLAHOMA STATE)",
    doi = "10.1016/0370-2693(87)91332-3",
    journal = "Phys. Lett. B",
    volume = "191",
    pages = "122--126",
    year = "1987"
}

@article{Stockdale:2025sxi,
    author = "Stockdale, Lucia and Volkas, Raymond R.",
    title = "{Higgs Yukawa coupling constraints on a benchmark one-loop radiative mass model for the bottom, charm and tau}",
    eprint = "2511.08924",
    archivePrefix = "arXiv",
    primaryClass = "hep-ph",
    month = "11",
    year = "2025"
}

@article{finetune,
    author = "Clarke, Jackson D. and Cox, Peter",
    title = "{Naturalness made easy: two-loop naturalness bounds on minimal SM extensions}",
    eprint = "1607.07446",
    archivePrefix = "arXiv",
    primaryClass = "hep-ph",
    doi = "10.1007/JHEP02(2017)129",
    journal = "JHEP",
    volume = "02",
    pages = "129",
    year = "2017"
}

@article{PTparams,
    author = "Peskin, Michael E. and Takeuchi, Tatsu",
    title = "{Estimation of oblique electroweak corrections}",
    reportNumber = "SLAC-PUB-5618",
    doi = "10.1103/PhysRevD.46.381",
    journal = "Phys. Rev. D",
    volume = "46",
    pages = "381--409",
    year = "1992"
}

@article{PTparamsearly,
    author = "Peskin, Michael E. and Takeuchi, Tatsu",
    title = "{A New constraint on a strongly interacting Higgs sector}",
    reportNumber = "SLAC-PUB-5272",
    doi = "10.1103/PhysRevLett.65.964",
    journal = "Phys. Rev. Lett.",
    volume = "65",
    pages = "964--967",
    year = "1990"
}

@article{kappaframe,
    author = "David, A. and Denner, A. and Duehrssen, M. and Grazzini, M. and Grojean, C. and Passarino, G. and Schumacher, M. and Spira, M. and Weiglein, G. and Zanetti, M.",
    collaboration = "LHC Higgs Cross Section Working Group",
    title = "{LHC HXSWG interim recommendations to explore the coupling structure of a Higgs-like particle}",
    journal = "",
    eprint = "1209.0040",
    archivePrefix = "arXiv",
    primaryClass = "hep-ph",
    reportNumber = "CERN-PH-TH-2012-284, LHCHXSWG-2012-001",
    month = "9",
    year = "2012"
}

@article{kappaframelong,
    author = "Andersen, J R and others",
    editor = "Heinemeyer, S and Mariotti, C and Passarino, G and Tanaka, R",
    collaboration = "LHC Higgs Cross Section Working Group",
    title = "{Handbook of LHC Higgs Cross Sections: 3. Higgs Properties}",
    journal = "",
    eprint = "1307.1347",
    archivePrefix = "arXiv",
    primaryClass = "hep-ph",
    reportNumber = "CERN-2013-004",
    doi = "10.5170/CERN-2013-004",
    month = "7",
    year = "2013"
}

@article{collider_future,
    author = "de Blas, J. and others",
    title = "{Higgs Boson Studies at Future Particle Colliders}",
    eprint = "1905.03764",
    archivePrefix = "arXiv",
    primaryClass = "hep-ph",
    reportNumber = "DESY-19-079",
    doi = "10.1007/JHEP01(2020)139",
    journal = "JHEP",
    volume = "01",
    pages = "139",
    year = "2020"
}

@article{STfuture,
    author = "de Blas, Jorge and Ciuchini, Marco and Franco, Enrico and Mishima, Satoshi and Pierini, Maurizio and Reina, Laura and Silvestrini, Luca",
    title = "{Electroweak precision observables and Higgs-boson signal strengths in the Standard Model and beyond: present and future}",
    eprint = "1608.01509",
    archivePrefix = "arXiv",
    primaryClass = "hep-ph",
    reportNumber = "KEK-TH-1919",
    doi = "10.1007/JHEP12(2016)135",
    journal = "JHEP",
    volume = "12",
    pages = "135",
    year = "2016"
}

@article{deBlas:2025gyz,
    author = "de Blas, Jorge and others",
    title = "{Physics Briefing Book: Input for the 2026 update of the European Strategy for Particle Physics}",
    eprint = "2511.03883",
    archivePrefix = "arXiv",
    primaryClass = "hep-ex",
    reportNumber = "CERN--2025-008, CERN-ESU-2025-001",
    doi = "10.23731/CYRM-2025-008",
    month = "11",
    year = "2025"
}

@article{radiative1,
    author = "Weinberg, Steven",
    title = "{Electromagnetic and weak masses}",
    doi = "10.1103/PhysRevLett.29.388",
    journal = "Phys. Rev. Lett.",
    volume = "29",
    pages = "388--392",
    year = "1972"
}

@article{radiative2,
    author = "Balakrishna, B. S. and Kagan, A. L. and Mohapatra, R. N.",
    title = "{Quark Mixings and Mass Hierarchy From Radiative Corrections}",
    reportNumber = "MdDP-PP-88-142",
    doi = "10.1016/0370-2693(88)91676-0",
    journal = "Phys. Lett. B",
    volume = "205",
    pages = "345--352",
    year = "1988"
}

@article{radiative3,
    author = "Babu, K. S. and Ma, Ernest",
    title = "{Radiative Mechanisms for Generating Quark and Lepton Masses: Some Recent Developments}",
    reportNumber = "MdDP-PP-89-193, UCRHEP-T37",
    doi = "10.1142/S0217732389002239",
    journal = "Mod. Phys. Lett. A",
    volume = "4",
    pages = "1975",
    year = "1989"
}

@article{radiative5,
    author = "Ma, Ernest",
    title = "{Hierarchical Radiative Quark and Lepton Mass Matrices}",
    reportNumber = "UCRHEP-T53",
    doi = "10.1103/PhysRevLett.64.2866",
    journal = "Phys. Rev. Lett.",
    volume = "64",
    pages = "2866--2869",
    year = "1990"
}

@article{radiative4,
    author = "He, Xiao-Gang and Volkas, Raymond R. and Wu, Dan-Di",
    title = "{Radiative Generation of Quark and Lepton Mass Hierarchies From a Top Quark Mass Seed}",
    reportNumber = "UM-P-89/58, OZ-P-89/21",
    doi = "10.1103/PhysRevD.41.1630",
    journal = "Phys. Rev. D",
    volume = "41",
    pages = "1630",
    year = "1990"
}

@article{radiative6,
    author = "Dobrescu, Bogdan A. and Fox, Patrick J.",
    title = "{Quark and lepton masses from top loops}",
    eprint = "0805.0822",
    archivePrefix = "arXiv",
    primaryClass = "hep-ph",
    reportNumber = "FERMILAB-PUB-08-049-T",
    doi = "10.1088/1126-6708/2008/08/100",
    journal = "JHEP",
    volume = "08",
    pages = "100",
    year = "2008"
}

@article{radiative7,
    author = "Ma, Ernest",
    title = "{Radiative Origin of All Quark and Lepton Masses through Dark Matter with Flavor Symmetry}",
    eprint = "1311.3213",
    archivePrefix = "arXiv",
    primaryClass = "hep-ph",
    reportNumber = "UCRHEP-T538-(NOV-2013)",
    doi = "10.1103/PhysRevLett.112.091801",
    journal = "Phys. Rev. Lett.",
    volume = "112",
    pages = "091801",
    year = "2014"
}

@article{radiative8,
    author = "C\'arcamo Hern\'andez, A. E. and Kovalenko, Sergey and Schmidt, Ivan",
    title = "{Radiatively generated hierarchy of lepton and quark masses}",
    eprint = "1611.09797",
    archivePrefix = "arXiv",
    primaryClass = "hep-ph",
    doi = "10.1007/JHEP02(2017)125",
    journal = "JHEP",
    volume = "02",
    pages = "125",
    year = "2017"
}

@article{radiative9,
    author = "Arbel\'aez, Carolina and C\'arcamo Hern\'andez, A. E. and Cepedello, Ricardo and Kovalenko, Sergey and Schmidt, Ivan",
    title = "{Sequentially loop suppressed fermion masses from a single discrete symmetry}",
    eprint = "1911.02033",
    archivePrefix = "arXiv",
    primaryClass = "hep-ph",
    doi = "10.1007/JHEP06(2020)043",
    journal = "JHEP",
    volume = "06",
    pages = "043",
    year = "2020"
}

@article{radiative10,
    author = "Weinberg, Steven",
    title = "{Models of Lepton and Quark Masses}",
    eprint = "2001.06582",
    archivePrefix = "arXiv",
    primaryClass = "hep-th",
    reportNumber = "UTTG-10-19",
    doi = "10.1103/PhysRevD.101.035020",
    journal = "Phys. Rev. D",
    volume = "101",
    number = "3",
    pages = "035020",
    year = "2020"
}

@article{radiative11,
    author = "Mohanta, Gurucharan and Patel, Ketan M.",
    title = "{Radiatively generated fermion mass hierarchy from flavor nonuniversal gauge symmetries}",
    eprint = "2207.10407",
    archivePrefix = "arXiv",
    primaryClass = "hep-ph",
    doi = "10.1103/PhysRevD.106.075020",
    journal = "Phys. Rev. D",
    volume = "106",
    number = "7",
    pages = "075020",
    year = "2022"
}

@article{radiative12,
    author = "Bonilla, Cesar and Carcamo Hernandez, A. E. and Kovalenko, Sergey and Lee, H. and Pasechnik, R. and Schmidt, Ivan",
    title = "{Fermion mass hierarchy in an extended left-right symmetric model}",
    eprint = "2305.11967",
    archivePrefix = "arXiv",
    primaryClass = "hep-ph",
    doi = "10.1007/JHEP12(2023)075",
    journal = "JHEP",
    volume = "12",
    pages = "075",
    year = "2023"
}

@article{radiative13,
    author = "Jana, Sudip and Klett, Sophie and Lindner, Manfred and Mohapatra, Rabindra N.",
    title = "{Radiative origin of fermion mass hierarchy in left-right symmetric theory}",
    eprint = "2409.04246",
    archivePrefix = "arXiv",
    primaryClass = "hep-ph",
    doi = "10.1007/JHEP01(2025)082",
    journal = "JHEP",
    volume = "01",
    pages = "082",
    year = "2025"
}

@article{radiative14,
    author = "Baker, Michael J. and Cox, Peter and Volkas, Raymond R.",
    title = "{Radiative muon mass models and $(g-2)_\mu$}",
    eprint = "2103.13401",
    archivePrefix = "arXiv",
    primaryClass = "hep-ph",
    doi = "10.1007/JHEP05(2021)174",
    journal = "JHEP",
    volume = "05",
    pages = "174",
    year = "2021"
}

@article{radiative15,
    author = "Mohanta, Gurucharan and Patel, Ketan M.",
    title = "{Partially flavour non-universal U(1) and radiative fermion masses}",
    eprint = "2508.05439",
    archivePrefix = "arXiv",
    primaryClass = "hep-ph",
    doi = "10.1007/JHEP02(2026)170",
    journal = "JHEP",
    volume = "02",
    pages = "170",
    year = "2026"
}

@article{ESUbriefing,
    author = "Ellis, Richard Keith and others",
    title = "Physics Briefing Book: Input for the European Strategy for Particle Physics Update 2020",
    journal = "",
    eprint = "1910.11775",
    archivePrefix = "arXiv",
    primaryClass = "hep-ex",
    reportNumber = "CERN-ESU-004",
    month = "10",
    year = "2019"
}

@article{CMS_cc,
    author = "Tumasyan, Armen and others",
    collaboration = "CMS",
    title = "{Search for Higgs Boson Decay to a Charm Quark-Antiquark Pair in Proton-Proton Collisions at s=13\,\,TeV}",
    eprint = "2205.05550",
    archivePrefix = "arXiv",
    primaryClass = "hep-ex",
    reportNumber = "CMS-HIG-21-008, CERN-EP-2022-081",
    doi = "10.1103/PhysRevLett.131.061801",
    journal = "Phys. Rev. Lett.",
    volume = "131",
    number = "6",
    pages = "061801",
    year = "2023"
}

@article{CMS_sum,
    author = "Tumasyan, Armen and others",
    collaboration = "CMS",
    title = "{A portrait of the Higgs boson by the CMS experiment ten years after the discovery.}",
    eprint = "2207.00043",
    archivePrefix = "arXiv",
    primaryClass = "hep-ex",
    reportNumber = "CMS-HIG-22-001, CERN-EP-2022-039",
    doi = "10.1038/s41586-022-04892-x",
    journal = "Nature",
    volume = "607",
    number = "7917",
    pages = "60--68",
    year = "2022",
    note = "[Erratum: Nature 623, (2023)]"
}

@article{ATLAS_bb,
    author = "Aad, Georges and others",
    collaboration = "ATLAS",
    title = "{Measurements of $WH$ and $ZH$ production in the $H \rightarrow b\bar{b}$ decay channel in $pp$ collisions at 13 TeV with the ATLAS detector}",
    eprint = "2007.02873",
    archivePrefix = "arXiv",
    primaryClass = "hep-ex",
    reportNumber = "CERN-EP-2020-087",
    doi = "10.1140/epjc/s10052-020-08677-2",
    journal = "Eur. Phys. J. C",
    volume = "81",
    number = "2",
    pages = "178",
    year = "2021"
}

@article{ATLAS_tau,
    author = "Aaboud, Morad and others",
    collaboration = "ATLAS",
    title = "{Cross-section measurements of the Higgs boson decaying into a pair of $\tau$-leptons in proton-proton collisions at $\sqrt{s}=13$ TeV with the ATLAS detector}",
    eprint = "1811.08856",
    archivePrefix = "arXiv",
    primaryClass = "hep-ex",
    reportNumber = "CERN-EP-2018-232",
    doi = "10.1103/PhysRevD.99.072001",
    journal = "Phys. Rev. D",
    volume = "99",
    pages = "072001",
    year = "2019"
}

@article{ATLAS_Z,
    author = "Aad, Georges and others",
    collaboration = "ATLAS",
    title = "{Higgs boson production cross-section measurements and their EFT interpretation in the $4\ell $ decay channel at $\sqrt{s}=$13 TeV with the ATLAS detector}",
    eprint = "2004.03447",
    archivePrefix = "arXiv",
    primaryClass = "hep-ex",
    reportNumber = "CERN-EP-2020-034",
    doi = "10.1140/epjc/s10052-020-8227-9",
    journal = "Eur. Phys. J. C",
    volume = "80",
    number = "10",
    pages = "957",
    year = "2020",
    note = "[Erratum: Eur.Phys.J.C 81, 29 (2021), Erratum: Eur.Phys.J.C 81, 398 (2021)]"
}

@article{ATLAS_gamma,
    author = "Aad, Georges and others",
    collaboration = "ATLAS",
    title = "{Measurement of the properties of Higgs boson production at $\sqrt{s} = 13$ TeV in the $H\to\gamma\gamma$ channel using $139$ fb$^{-1}$ of $pp$ collision data with the ATLAS experiment}",
    eprint = "2207.00348",
    archivePrefix = "arXiv",
    primaryClass = "hep-ex",
    reportNumber = "CERN-EP-2022-094",
    doi = "10.1007/JHEP07(2023)088",
    journal = "JHEP",
    volume = "07",
    pages = "088",
    year = "2023"
}

@article{SUSYbound,
    author = "Aaboud, Morad and others",
    collaboration = "ATLAS",
    title = "{Search for heavy charged long-lived particles in the ATLAS detector in 36.1 fb$^{-1}$ of proton-proton collision data at $\sqrt{s} = 13$ TeV}",
    eprint = "1902.01636",
    archivePrefix = "arXiv",
    primaryClass = "hep-ex",
    reportNumber = "CERN-EP-2018-339",
    doi = "10.1103/PhysRevD.99.092007",
    journal = "Phys. Rev. D",
    volume = "99",
    number = "9",
    pages = "092007",
    year = "2019"
}

@article{ParticleDataGroup:2026aaa,
    author = "Takahashi, F. and others",
    collaboration = "Particle Data Group",
    title = "{Review of Particle Physics}",
    doi = "10.1142/S0217751X26300115",
    journal = "Int. J. Mod. Phys. A",
    volume = "41",
    pages = "2630011",
    year = "2026"
}

@PREAMBLE{
 "\providecommand{\noopsort}[1]{}" 
 # "\providecommand{\singleletter}[1]{#1}%" 
}

\end{document}